\documentclass[aps,twocolumn,prd,superscriptaddress,nofootinbib,floatfix]{revtex4-2}

\usepackage{mathtools}
\usepackage{amsfonts}
\usepackage{amssymb}
\usepackage{mathrsfs}
\usepackage{bbm}
\usepackage{slashed}
\usepackage{amsmath}
\usepackage{bm}

\usepackage{graphicx}
\usepackage{color}
\usepackage{colortbl}
\usepackage{array}

\usepackage{float}
\usepackage{placeins}
\usepackage{booktabs}
\usepackage[caption=false]{subfig}
\usepackage{makecell}
\usepackage{tabstackengine}

\usepackage{xspace}
\usepackage{hyperref}
\usepackage[nameinlink]{cleveref}
\usepackage{bookmark}
\usepackage{siunitx}

\usepackage{xifthen}
\usepackage{xcolor}
\hypersetup{
	colorlinks,
	linkcolor={red!75!black},
	citecolor={blue!75!black},
	urlcolor={blue!75!black}
}

\usepackage[utf8]{inputenc}

\newcommand{\eb}{\epsilon_{{b}}}
\newcommand{\muq}{\mu_{\textrm{q}}}

\newcommand{\mub}{\mu_{{b}}}

\newcommand{\MeV}{\;\text{MeV}}
\newcommand{\GeV}{\;\text{GeV}}
\newcommand{\fm}{\;\text{fm}}

\newcommand{\tr}{\text{tr}}

\newcommand{\feyn}[1]{
	\setbox0=\hbox{\ensuremath{#1}}
	\hbox to\wd0{\hbox to0pt{\hbox to\wd0{\hss/\hss}\hss}\box0}}

\newcommand{\nsat}{n_{b}^{\text{(sat)}}}
\newcommand{\Vb}{V_{b}}
\newcommand{\nb}{n_{b}}

\def\pslash{p\hspace{.015cm}\llap{/}\hspace{.015cm}}

\setkeys{Gin}{width=0.48\textwidth}
\newcommand{\imag}{\text{i}}

\graphicspath{{./figures/}}

\newcommand{\gettitle}{Binding Energy}

\newcommand{\getHeidelbergAffiliation}{\affiliation{Institut f\"{u}r Theoretische Physik, Universit\"{a}t Heidelberg, Philosophenweg 16, 69120 Heidelberg, Germany}}
\newcommand{\getTokyoAffiliation}{\affiliation{Department of Physics, The University of Tokyo, 
7-3-1 Hongo, Bunkyo-ku, Tokyo 113-0033, Japan}}
\newcommand{\getEMMIAffiliation}{\affiliation{ExtreMe Matter Institute EMMI, GSI, Planckstr. 1, 64291 Darmstadt, Germany}}
\newcommand{\getCologneAffiliation}{\affiliation{Institut f\"ur Theoretische Physik, Universit\"at zu K\"oln, 50937 Cologne, Germany}}
\newcommand{\getDarmstadtAffiliation}{\affiliation{Institut für Kernphysik, Technische Universität Darmstadt, D-64289 Darmstadt, Germany}}

\hypersetup{
	colorlinks,
	linkcolor={red!75!black},
	citecolor={blue!75!black},
	urlcolor={blue!75!black},
	pdftitle={\gettitle},
	pdfauthor={Zelle},
	pdfkeywords={effective theory} {analytic continuation}
	{correlations functions} {hadronization}
	{functional renormalisation group}
	{real time} {bound states}
	bookmarksopen=true,
	bookmarksopenlevel=2,
	bookmarksnumbered=true
}

\begin{document}

	\title{The nuclear liquid-gas transition in QCD}
	
	\author{Kenji Fukushima} \getTokyoAffiliation
	\author{Jan Horak} \getHeidelbergAffiliation
	\author{Jan M.~Pawlowski} \getHeidelbergAffiliation \getEMMIAffiliation
	\author{Nicolas Wink} \getDarmstadtAffiliation
	\author{Carl Philipp Zelle}\getCologneAffiliation

\begin{abstract} 
We estimate the nuclear saturation density and the binding energy in a nuclear liquid from precision data on the coupling of the four-quark scattering vertex in the vector channel, computed within functional QCD. We show that this coupling is directly related to the density-density potential and the latter is used for the estimates. In a first qualitative computation we find a saturation density of 0.2\,fm${}^{-3}$ and an upper bound for the binding energy of 21.5\,MeV, in agreement with the empirical values of 0.16\,fm${}^{-3}$  and 16\,MeV, respectively. We also use the scattering vertex for constructing an emergent low-energy effective theory for the liquid gas transition from QCD correlation function, whose coupling parameters can be determined within QCD. As a first consistency check of this construction we estimate the in-medium reduction of the nucleon pole mass. 
\end{abstract}
	
\maketitle

\section{Introduction}

One of the few experimentally accessible aspects of the low-temperature phase structure of strongly interacting matter is the liquid-gas transition in symmetric nuclear matter~\cite{Bertsch:1983uv}. It is a first-order phase transition at which nucleons pile up into a nuclear liquid, and nuclei can be  thought of as droplets of this liquid. The binding energy in the infinite nuclear liquid is known to be $\SI{16}{\MeV}$ at the saturation density $n_{b}=\nsat\simeq 0.16{\fm^{-3}}$ at vanishing temperature. In the nuclear experiment finite-size effects are crucial but some signatures have been confirmed; see \cite{Chomaz:2002xxf} for a summary of measured observables.

This phenomenon is less understood from first principles theory, i.e., quantum chromodynamics (QCD), as it combines a couple of challenging problems of strongly correlated low-energy QCD. The understanding of the liquid-gas transition necessitates the determination of the nucleon pole mass $m_{b}$ and the excitation energy per nucleon $E/A$ as a function of the baryon density $n_{b}$. The minimum location and the depth of $E/A$ give the saturation density $\nsat$ and the binding energy $\epsilon_{b}$, respectively. The determination of the nucleon mass in the vacuum has been achieved from Dyson-Schwinger (DSE)--Bethe-Salpeter (BSE)--Faddeev equations; see \cite{Eichmann:2016yit} for a recent review. This result has been reproduced within lattice-QCD simulations; see \cite{Brambilla:2014jmp} for a review. As for the other two observables, the saturation density $\nsat$, and the binding energy $\epsilon_{b}$, the energy density functional has not been derived yet from first principles QCD except in special regimes; see e.g.~\cite{deForcrand:2009dh,Kim:2023dnq} for lattice-QCD approaches to nuclear matter at strong coupling.

As long as the baryon density is close to the saturation density, the protocol of \textit{ab initio} calculations of QCD within a derivative expansion ought to be valid.  In this way the nuclear equation of state (EOS) has been estimated for symmetric and asymmetric nuclear matter; see \cite{Holt:2013fwa} for a review of pioneering works.
The chiral approach continues up to next-to-next-to-next-leading order (N$^3$LO) including many-body interactions \cite{Drischler:2017wtt}, which has been applied to constrain properties of neutron stars \cite{Drischler:2020fvz}.
Although the chiral effective theory has predictive power near the saturation density,
the validity region is limited below the baryon density $1.2$-$1.5 \, \nsat$.
Thus, it is desirable to complement our understanding of the liquid-gas transition of nuclear matter with microscopic descriptions with a more direct connection to QCD.
In this direction a hybrid method using two complementary approaches has been discussed; see \cite{Leonhardt:2019fua}.

In the present work we demonstrate that the four-quark scattering amplitude $\lambda_\omega(\vec p)$ in the density channel sets the relevant scale for the liquid-gas transition. This scattering amplitude has been computed in a first principles functional renormalisation group approach to QCD \cite{Mitter:2014wpa, Rennecke:2015eba, Cyrol:2017ewj}, The amplitude $\lambda_\omega(\vec p)$ or its Fourier transform in position space exhibits a repulsive short-distance part for $r\to 0$ and an attractive long-distance part for $r\to \infty$. We translate $\lambda_\omega(\vec p)$ into a potential energy of density correlations, and the stable point leads to a direct estimate of the saturation density as well as the binding energy.  Interestingly, $\nsat$ and $\epsilon_{b}$ deduced from the potential energy turn out to be compatible with the empirical values.

The first principles QCD correlation functions in \cite{Mitter:2014wpa, Braun:2014ata, Rennecke:2015eba, Cyrol:2017ewj, Fu:2019hdw} can also be used to compute the emergent low-energy effective theory of baryons, mesons, and quarks along the lines set up in \cite{Fukushima:2021ctq}. This low-energy effective theory of QCD  governs the dynamics of the liquid-gas transition of nuclear matter, and unlike conventional model studies its parameters are fixed by the QCD results in \cite{Mitter:2014wpa, Rennecke:2015eba, Cyrol:2017ewj}. For instance, $\lambda_\omega$ has been a fitting parameter in phenomenological approaches such as quantum hadrodynamics (QHD) by Serot and Walecka \cite{Serot:1997xg}. The covariantly extended formalism of relativistic mean-field nuclear calculations still needs this parameter $\lambda_\omega$, which is usually inferred from the intrinsic property of the $\omega$ meson. In contradistinction, in the present work we utilise the momentum dependence of $\lambda_\omega(\vec p)$ computed in QCD. This momentum dependence encodes rich physics, which is not fully taken into account in simple mean-field calculations. We note that this momentum dependence is reminiscent of the effective nuclear potential, such as the Skyrme interaction and the Gogny interaction, that is density-dependent, for a recent review, see \cite{Shen:2019dls}. This can be understood from the correspondence between the density and the Fermi momentum, or that between the characteristic inter-particle distance at a certain density and the momentum scale. In this sense first principles data on $\lambda_\omega(\vec p)$ sheds light on the density-dependent phenomenological potential in nuclear physics.

\section{The nuclear liquid from density-density correlations}
\label{sec:BindingEnergyDDPot}

\begin{figure}[t]
	\centering
	\includegraphics[width=.95\linewidth]{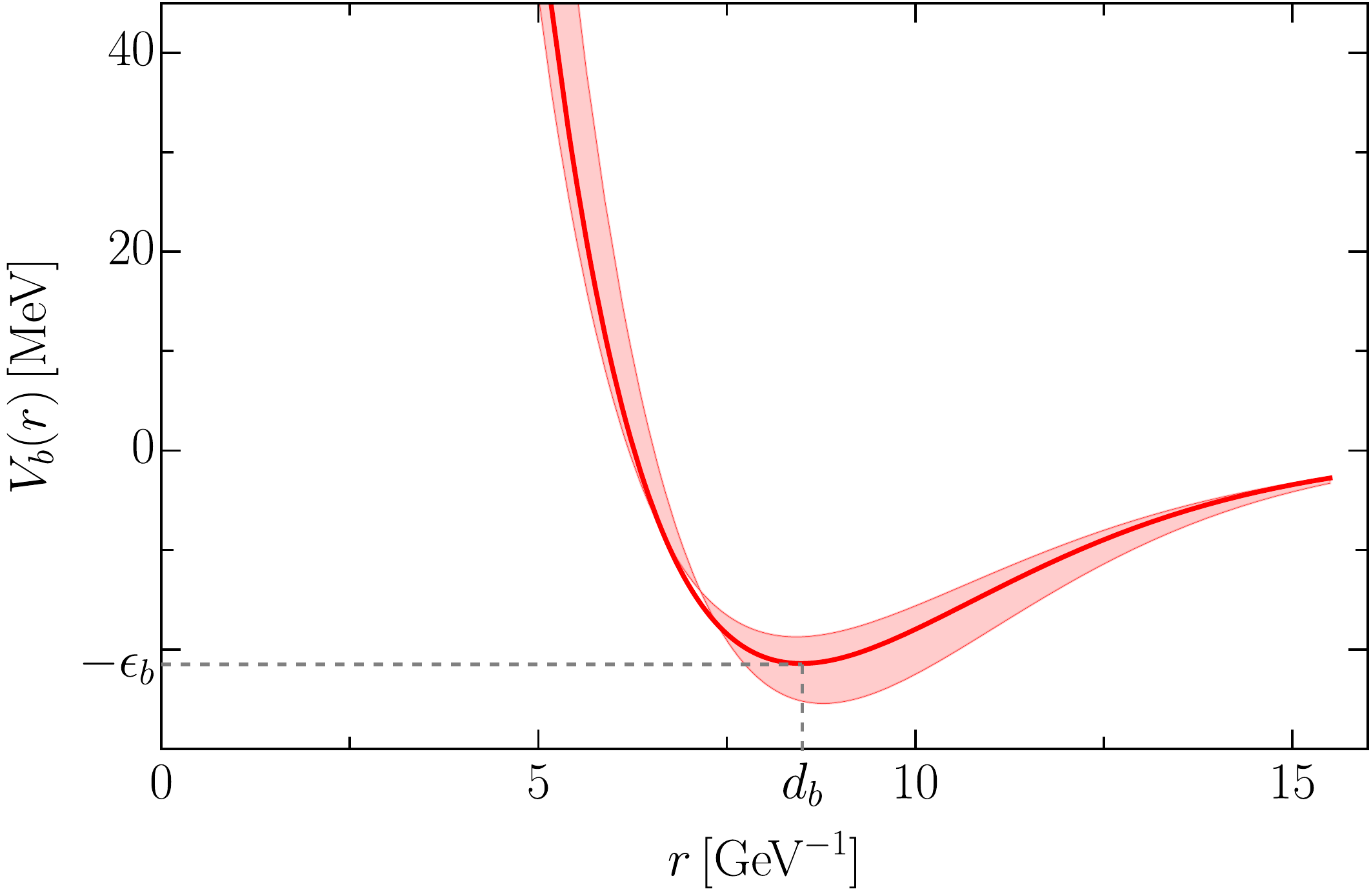}
	\caption{Static density-density potential $\Vb(r)$~\labelcref{eq:VbFourier} as a function of distance $r$ in the nuclear liquid, computed from QCD correlation functions in \cite{Cyrol:2017ewj}. We identify the position of the minimum of the potential, $d_b$, with a typical length scale for the distance between nucleons in the nuclear liquid, and the depth of the minimum with the binding energy $\eb$. These are found to be $d_b \approx \SI{8.5 \pm 0.3}{\GeV^{-1}}$ and $\epsilon_b \approx\SI{21\pm5}{MeV}$.}
	\label{fig:VectorPotentialX}
\end{figure}

In this section we put forward an approach for computing the binding energy $\eb$ in a nuclear liquid at the saturation density $\nsat$ from quark and baryon correlation functions in first principles QCD. These observables are encoded in the static potential between nucleons or the density-density potential $V_b(r)$ in the liquid at a distance $r$, and its estimate from QCD correlation functions is presented in \Cref{fig:VectorPotentialX}. In particular, the density is obtained from the location $d_b$ of the minimum of the potential, as $d_b$ is related to the distance between nucleons in the nuclear liquid. The  depth of the potential at $d_b$ offers an estimate for the binding energy $\eb$.
%

\subsection{Density-density potential in the nuclear liquid}
\label{sec:Vb}

We proceed with extracting the density-density potential $\Vb(r)$ in the nuclear liquid from density-density correlations in first principles QCD: the static baryon density $\nb(\vec x)$ with $\partial_t \nb(x)=0$ describes a number of nucleons in ${\cal N}_x$,
\begin{align} 
	\int_{\vec x\in {\cal N}_x} \nb(x)\,,
	\label{eq:nb}
\end{align}
where ${\cal N}_x$ is some infinitesimal neighbourhood of $\vec x$. In the present spatially homogeneous situation, \labelcref{eq:nb} is directly related to the average density, the first density moment, computed by a $\mu_b$-derivative of the grand potential $\Omega$ in QCD, 
\begin{align}
	\nb = \frac{1}{{\cal V}_4} \int_x \langle \hat n_b(x)\rangle = -\frac{1}{{\cal V}_4 }\frac{\partial \Omega  }{\partial \mub}\,, 
	\label{eq:IntDensity} 
\end{align}
where we have divided out the space-time volume ${\cal V}_4$. The grand potential $\Omega$ is nothing but the effective action $\Gamma_{\textrm{QCD}}$ of QCD, evaluated on the equations of motion (EoM), i.e., 
\begin{align}
	\Omega =\left. \Gamma_\textrm{QCD}\right|_\textrm{EoM}\,.
\end{align}
For $\mu_b<\mu_b^{\textrm{os}}$, the density is vanishing. Here, $\mu_b^{\textrm{os}}$ is the onset chemical potential.  For larger baryon chemical potentials, $\mu_b>\mu_b^\textrm{os}$, the density is non-vanishing. At the onset chemical potential for symmetric nuclear matter, the density jumps to the saturation density 
\begin{align}
	\nsat= \nb( \mub^\text{(os)})\neq 0\,,
\end{align}
at the first order nuclear liquid-gas transition. In the present work we estimate the size of $\nsat$ from $\Vb(r)$, but in \Cref{sec:BindingEnergyLEFT} we also discuss its derivation from effective field theory considerations. 

The two-body potential $\Vb(r)$ is the second static moment of local density fluctuations in position space and is given by 
\begin{align} 
	\int_{x,y}\langle \hat n_b(x) \hat n_b(y)\rangle _{\mathrm{c}}  =  \frac{\partial^2 \Omega}{\partial \mub^2}\,,
\label{eq:Gnbnb}
\end{align}
where the subscript c indicates that the second derivative of the grand potential w.r.t.~the baryon chemical potential provides the connected part of the density-density correlation, 
\begin{align}
	\langle \hat n_b(x)  \hat n_b(y)\rangle_{\mathrm{c}}  =  \langle  \hat n_b(\vec x)  \hat n_b(\vec y)\rangle -n_b(\vec x) n_b(\vec y) \,. 
\label{eq:nbnbConnected}
\end{align}
The integrand \labelcref{eq:nbnbConnected} of \labelcref{eq:Gnbnb} comprises both observables under investigation: $\nsat$ and $\epsilon_b$.

It is left to compute the density correlation \labelcref{eq:Gnbnb} from correlation functions in first principles QCD. While \labelcref{eq:Gnbnb} is the integrated correlation, we are interested in the local one, \labelcref{eq:nbnbConnected}, 
\begin{align}
	\langle \hat n_b(x) \hat n_b(y)\rangle_c=\frac{\delta^2 \Gamma_\textrm{QCD}}{\delta \mu_b(x) \delta \mu_b(y)}\,, 
	\label{eq:GnbnbLocal}
\end{align}
evaluated for a static situation. Note that in contrast to~\labelcref{eq:Gnbnb}, the derivatives in~\labelcref{eq:GnbnbLocal} are functional derivatives, which compensate for the two spacetime integrations in~\labelcref{eq:Gnbnb} by giving rise to delta distributions. 

The respective term in the QCD effective action, formulated in terms of the baryon density, is given by 
\begin{align}
	\Gamma_{n_b n_b} = - \int_{x,y} n_b(x) \,\lambda_{n_b n_b}(x-y)\, n_b(y)\,,
	\label{eq:Gammanbnb}
\end{align}
with the interaction strength $ \lambda_{n_b n_b}(x-y)$. We emphasise that 	\labelcref{eq:Gammanbnb} strictly holds in a density functional approach to QCD, where the baryon or quark density has been introduced as a dynamical variable. This is tantamount to introducing a dynamical field for the vector mode $\omega_0$ via dynamical hadronisation as discussed in \cite{Fukushima:2021ctq}. 

The static equal time part of \labelcref{eq:Gammanbnb} is obtained for $\partial_t n_b(t,\vec x)=0$. Inserting such a static density distribution into \labelcref{eq:Gammanbnb} leads us to 
\begin{align}
	\frac{1}{T}	\Gamma_{n_b n_b} = 	- \int_{\vec x,\vec y} n_b(\vec x) V_b(r)  n_b(\vec y) \,,
	\label{eq:GammanbnbStatic}
\end{align}
where the potential $V_b(r)$ is nothing but the dressing factor of the density-density term and $r$ is the distance between the density locations, 
\begin{align}
	\qquad r=\|\vec x-\vec y\|\,. 
	\label{eq:Distance} 
\end{align}
In \labelcref{eq:GammanbnbStatic} we have divided out the time extent $T$ with the space-time volume ${\cal V}_4 = T\,{\cal V}_3$, where ${\cal V}_3$ is the spatial volume. The density-density potential $V_b(r)$ in \labelcref{eq:GammanbnbStatic} is simply given by the static part of the coupling $\lambda_{n_b n_b}$, 
\begin{align}
	V_b(r) = \int_{-T/2}^{T/2} dt\, \lambda_{n_b n_b}(t,r)\,.
	\label{eq:VbStaticlambda}
\end{align}
The potential $V_b(r)$ in \labelcref{eq:VbStaticlambda} can be obtained as Fourier transform from momentum dependent data for $\lambda_{n_b n_b}(p)$ via 
\begin{subequations}
\begin{align}
	V_b(r) \simeq 	\int \frac{\textrm{d}^3 p}{(2\pi)^3} V_b(\vec p)\, e^{\imag \vec p(\vec x-\vec y)} \,, 
\end{align}
where we identify
\begin{align}
	V_b(\vec p) = \lambda_{n_b n_b}(p_0=0, \vec p) \,.
\end{align}
\label{eq:VbFourier}
\end{subequations}

In \Cref{fig:VectorPotentialX}, we display our estimate for the potential $V_b(r)$ \labelcref{eq:VbFourier} of static density-density correlations obtained from fundamental QCD correlation functions in \cite{Cyrol:2017ewj}. The distance $r=d_b$ of its minimum is the distance of ``cells'' in the gas or fluid and hence is directly related to the density. In turn, its depth $-V_b(d_b)$ provides an estimate for the binding energy.

\subsection{Density-density correlations from QCD}
\label{sec:DensityDensityQCD}
 
In the present work we utilise QCD correlation functions from \cite{Cyrol:2017ewj}, computed in terms of quark, gluons and mesons, and hence we rewrite \labelcref{eq:Gammanbnb} as a correlation of the quark-density with $n_q= 3 n_b$, 
\begin{align}
	\Gamma_{n_b n_b} = 	-\frac19  \int_{x, y} n_q( x)\, \lambda_{n_b n_b}(x-y)\,  n_q(y) \,. 
	\label{eq:Gammanqnq}
\end{align}
Hence, the coupling strength of the quark density is a ninth of that of the baryon density. The density operator $n_q$ is proportional to the quark bilinear $\bar q\gamma_0 q$, but also features the wave function $Z_q(p)$ of the quark, 
\begin{align}
	n_q\propto Z_q \bar q\gamma_0 q\,.
	\label{eq:densityDefZ}
\end{align}
The wave function is the ``dressing'' of the full kinetic term of the QCD effective action and reads 
\begin{align}
	\Gamma_{q,\textrm{kin}} = \int_p  Z_q(p)  \bar q(p) \bigg[ \imag \pslash -\gamma_0 \mu_q + M_q(p) \bigg] q(-p) \,, 
	\label{eq:Gammaqkin}
\end{align}
where we have dropped a further splitting of the tensor structure in the presence of the chemical potential, which singles out the rest frame. The wave function $Z_q(p)$ absorbs the RG-scaling of the quark fields, leaving us with an RG-invariant mass function $M_q(p)$. More generally, all fields in the effective action have to be accompanied by their respective wave functions $Z^{1/2}(p)$, leaving us with RG-invariant dressings such as $M_q$. 

The baryon density-density contribution to the effective action \labelcref{eq:Gammanqnq} in momentum space takes the form 
\begin{align}
	\Gamma_{n_b n_b} = 	- \frac19 \int_{p} n_q(-p) \,\lambda_{n_b n_b}(p)\, n_q(p) \,. 
	\label{eq:GammanqnqMomentum}
\end{align}
Taking into account~\labelcref{eq:densityDefZ}, in the lowest order approximation the density in momentum space $n_q(p)$ takes the form 
\begin{align}
	n_q(p)\simeq \int_l Z^{1/2}_q(l-p) Z^{1/2}_q(p)\bar q(l-p)\gamma_0 q(p)\,.
	\label{eq:nqBilinearMomentum}
\end{align}
\Cref{eq:nqBilinearMomentum} holds in the mean field or Gau\ss ian approximation, and its corrections contain higher order terms of the density quark bilinear and further higher order terms in the quarks and anti-quarks. This complete relation has been exhaustively studied at the example of the scalar-pseudoscalar channel: in \cite{Cyrol:2017ewj} and other works, the scalar-pseudoscalar channel has been treated within dynamical hadronisation, and the emergent scalar $\sigma$-mode and pions $\boldsymbol{\pi}^T=(\pi_1,\pi_2,\pi_3)$ have been introduced. To the lowest order, this amounts to identifying e.g.~$\sigma(x) \propto \bar q(x) q(x)$ by the equations of motion of $\sigma$. However, at the full quantum level higher order terms in the bilinear $\bar q(x) q(x)$ (and others) enter via the full effective potential $V_\textrm{eff}((\sigma^2 + \boldsymbol{\pi}^2)/2)$. These terms are relevant for the non-trivial pion dynamics in the infrared that generates chiral perturbation theory. 

In contradistinction to the important dynamics in the pseudoscalar channel, it has been shown in \cite{Fukushima:2021ctq}, that the identification \labelcref{eq:nqBilinearMomentum} holds  true for $\mu_q \leq \mu^\textrm{(os)}_q$. Note that the analysis in \cite{Fukushima:2021ctq} was done in terms of $\omega_0$, the zero component of the vector iso-scalar channel field $\omega_\mu$. In the following we assume the approximation \labelcref{eq:nqBilinearMomentum} to also hold for chemical potentials $\mu_q \gtrsim  \mu^\textrm{(os)}_q$ close the onset chemical potential. 

In conclusion, we can directly use the results in \cite{Cyrol:2017ewj} for the density channel of the four-quark scattering terms,  
\begin{align}\nonumber 
	\Gamma_{n_q n_q} \simeq&\,  -\frac14 \int_{p}	\lambda_\omega(p_1,p_2,p_3) \\[1ex] 
&\hspace{1cm}\times 	\left[\bar q(p_1) \gamma_0 q(p_2) \right] \left[\bar q(p_3) \gamma_0 q(p_4) \right]\,,
	\label{eq:GmatDensityChannel}
\end{align}
with $p=(p_1,p_2,p_3)$ and $p_4=-(p_1+p_2+p_3)$. In \labelcref{eq:GmatDensityChannel},  $\lambda_\omega$ is a combination of couplings $\lambda^{(V\pm A)}$  to vector--axial-vector four-quark operators as used in \cite{Mitter:2014wpa} and \cite{Cyrol:2017ewj}, 
\begin{align}
	\lambda_\omega = \lambda^{(V+A)} + \lambda^{(V-A)}\,.
\end{align}
In \cite{Cyrol:2017ewj}, the vertex dressing $\lambda_\omega(p_1,p_2,p_3)$ has been evaluated at the momentum $p$ for the anti-quarks and $-p$ for the quarks, 
\begin{align}
	\lambda_\omega(p) = \lambda_\omega(p,-p,p)\,.
	\label{eq:lambdaomegapp}
\end{align}	
This is precisely the momentum configuration of the quark number current $n_q$~\labelcref{eq:densityDefZ}, that couples to the chemical potential in~\labelcref{eq:Gammaqkin}. Hence, within a qualitative estimate we identify the RG-invariant coupling $\lambda_{n_b n_b}(p)$ with the coupling in the vector channel (up to a factor 9/4),  
\begin{align}
	\lambda_{n_b n_b}(p) \approx \frac{9}{4} \bar\lambda_\omega(p)\,,\qquad  \bar\lambda_\omega(p)=\frac{\lambda_\omega(p)}{Z_q(p)^2}\,, 
	\label{eq:lambdanb-lambdaomega}
\end{align}
as the RG-invariant coupling between two density operators \labelcref{eq:Gammaqkin}. In \labelcref{eq:lambdanb-lambdaomega} we have also introduced the RG-invariant vector coupling $\bar\lambda_\omega$, deduced from the vertex dressing $\lambda_\omega$ and the quark wave function $Z_q$. We emphasise that the vertex dressing $\lambda_\omega$ of the four-quark correlation function is not an RG-invariant quantity. 

\begin{figure}[t]
	\centering
	\includegraphics[width=0.95\linewidth]{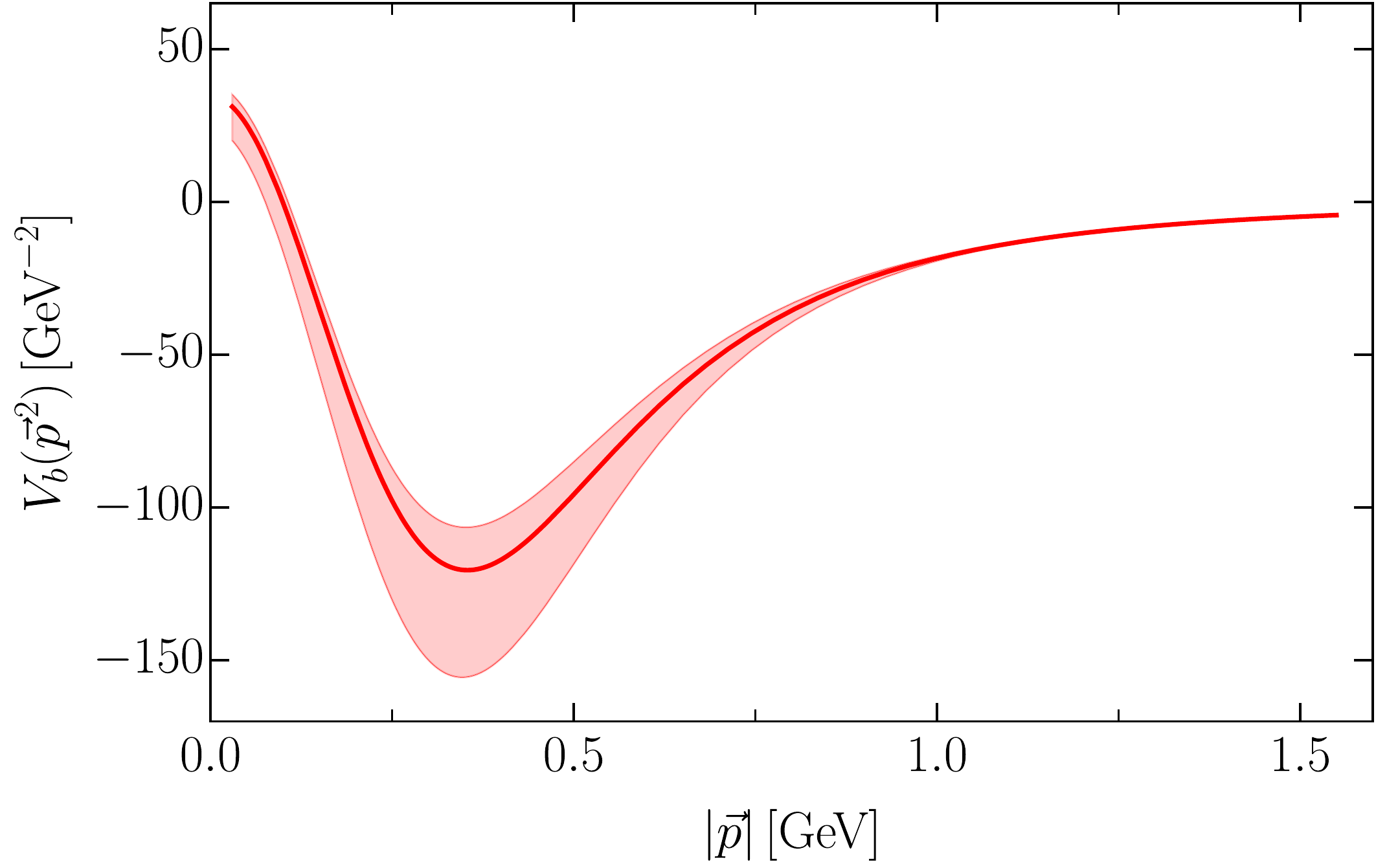}
	\caption{Static density-density potential in momentum space, see~\labelcref{eq:VbFourier,eq:lambdanb-lambdaomega}, using the $\bar\lambda_\omega$-data of~\cite{Cyrol:2017ewj}. A fit of $\bar\lambda_\omega$ is provided in \Cref{tab:FitPar} in \Cref{app:FitVb}. In the argument of $V_b(\vec p^{\,2})$ we made apparent that since the input data is obtained in the vacuum, no rest frame is singled out. The value of the potential in the minimum, \labelcref{eq:Minimumlambadnb}, is used for an estimate of the nucleon mass in the liquid, see \labelcref{eq:Mbliquid}.}
	\label{fig:VectorPotentialMom}
\end{figure}

These preparations allow us to compute $V_b(r)$ as the Fourier transform of the RG-invariant four-quark scattering coupling~\labelcref{eq:VbFourier}. The latter momentum space potential is displayed in \Cref{fig:VectorPotentialMom}, where we exploited that the data of~\cite{Cyrol:2017ewj} is obtained in the vacuum, and hence 
\begin{align}
	\lambda_\omega(p_0 = 0, \vec p) = \lambda_\omega(p^2 = \vec p^{\,2}) \,,
\end{align}
where $p^2=p_0^2+\vec p^{\,2}$ in our convention.
The Fourier transform of the momentum space potential leads to \Cref{fig:VectorPotentialX}. Hence, the error estimates in \Cref{fig:VectorPotentialMom,fig:VectorPotentialX} are that of $\lambda_\omega$ in \cite{Cyrol:2017ewj}, translated to the respective quantities here.  

%

\subsection{The nuclear liquid} 
\label{sec:nuclear-liquid}

We are now in the position to provide estimates for the onset density as well as the nucleon binding energy from first principles QCD. A last step concerns the total error estimate. Apart from the errors already mentioned there is a further one originating in the translation of the two-flavor scales to the physical case. In both cases the system is tuned to a pion mass $m_\pi\approx \SI{140}{\MeV}$. The off-shell dynamics of the strange quark influences the momentum dependence of $\lambda_\omega$, as it does the size and momentum-dependence of the light constituent quark mass 
\begin{align}
	M_l=M_l(0)\,, 
\label{eq:Mconst}
\end{align}
see e.g.~\cite{Fu:2019hdw}. This leads to a rescaling factor 
\begin{align} \label{eq:gamma_m}
	\gamma\approx \frac{M_l^{(2+1)} }{M_l^{(2)} } = 0.92\,, 
\end{align}
for momenta and $1/\gamma$ for distances of some observables, leaving us with roughly a further 10\% error. Our combined conservative error estimate sums up to 25\%, where a further 5\% span was added.

The effective nucleon potential $V_b$ in \labelcref{eq:VbFourier}, depicted in \Cref{fig:VectorPotentialX}, clearly displays a minimum at a distance $d_b$ with 
\begin{align}
	d_b = \SI{8.5 \pm 2.1}{\GeV^{-1}} \approx \SI{1.67\pm 0.42}{fm} \,, 
	\label{eq:min-dist}
\end{align}
with the 25\% error estimate. This allows us to provide an upper bound for the onset density: we estimate it by a cubic lattice density \mbox{$n_b \sim 1/d_b^3$}, which rather is the density of a solid state system. We obtain 
\begin{align}
	n^\text{(sat)}_{b}= \SI{0.21\pm 0.16}{\textrm{fm}^{-3}} \,, 
	\label{eq:density_lattice}
\end{align}
where we used Gau\ss ian error propagation of the error on $d_b$ in \labelcref{eq:min-dist}. 
This QCD-based estimate is in good agreement with the measured nuclear density of \mbox{$n^{\textrm{exp}}_{b} = \SI{0.16}{\fm^{-3}}$}. The potential $V_b$ also allows for an estimate or rather upper bound of the nucleon binding energy by the depth of the potential. This leads us to  
\begin{align}
	\epsilon_{b} \lesssim \SI{21 \pm 5}{MeV} \,. 
	\label{eq:binding-energy}
\end{align}
This bound is in good agreement with the experimentally established literature value for the nuclear binding energy of $\epsilon^\textrm{exp}_{b} = \SI{16}{\MeV}$ per nucleon.

In summary, the values of the estimates \labelcref{eq:density_lattice,eq:binding-energy} are very promising and support the validity of the current first principles approach. However, the large errors have to be systematically reduced in future applications, and we envisage work in two obvious directions: First, the density can be extracted from the full form of density-density potential $V_b$, a corresponding fit is provided in \Cref{app:FitVb}. 
Second, the estimates can be improved by a direct computation of the nucleon-nucleon potential, which can be done in terms of the functional renormalisation group approach with emergent hadrons and is described in \Cref{sec:BindingEnergyLEFT}. However, both extensions are beyond the scope of the present work.

\section{The nuclear liquid from emergent hadrons}
\label{sec:BindingEnergyLEFT} 

The results in \Cref{sec:BindingEnergyDDPot} for the saturation density and the binding energy in a nuclear liquid have been derived from the density-density term in the QCD effective action \labelcref{eq:GammanbnbStatic}, whose coupling was inferred from the four-quark scattering in the density channel of QCD. It is therefore suggestive to introduce the density mode and other emergent composites as dynamical degrees of freedom, hence facilitating the direct access to observables in low-energy QCD. This simple idea has been used very successfully since many decades within low-energy effective theories of QCD. If embedded directly in QCD, this finally allows for a direct computation of the nucleon-nucleon potential and hence gives a direct access to the physics of the liquid gas transition also beyond the few but crucial observables estimated here. 
 
In the functional renormalisation group approach these low-energy effective theories of QCD and the respective composite degrees of freedom emerge dynamically within first principles QCD. This allows us to determine the couplings such as $\lambda_\omega$ in the low-energy effective theories from that of first principle QCD correlation functions.

\subsection{Low-Energy effective theory with emergent vector mesons}
\label{sec:EmergentVectors}

So far this formulation has mostly been used for the scalar-pseudoscalar channel in QCD \cite{Gies:2002hq, Braun:2009ewx, Mitter:2014wpa, Braun:2014ata, Rennecke:2015eba, Cyrol:2017ewj, Fu:2019hdw, Fukushima:2021ctq} and the vector channel \cite{Rennecke:2015eba}; for reviews see \cite{Dupuis:2020fhh, Fu:2022gou}. In \cite{Fukushima:2021ctq} the general set-up with emergent composites has been described, including also baryons.  For the four-quark interactions this amounts to a scale-dependent Hubbard-Stratonovich transformation whose scale-dependence generates also higher order terms and makes the transformation exact. For the present purpose this is done by rewriting the four-quark interaction term \labelcref{eq:GmatDensityChannel} in the density channel in terms of the density mode $\omega_0$. This leads us to 
\begin{align}
	\Gamma_{n_q n_q}  \simeq &  \Biggl[- \int_p  \bar q(-p)\gamma_0\left( \frac{ h_\omega(p)}{2} \omega_0\right)   \,q(p) \\[1ex] \nonumber
	& \hspace{-.6cm}+
	\frac12\int_p \, Z_\omega(p)\,\omega_0(-p) \left[ p^2 + m_\omega^2 \right]\,\omega_0(p) \Biggr]_{\omega_0=\omega_0^\text{EoM}}\,, 
	\label{eq:Gomega}
\end{align}
with the constant solution of the $\omega_0$-equation of motion,   
\begin{align}
	\omega^\text{EoM}_0= \frac{1}{{\cal V}_4}\int_p  \frac{1}{Z_\omega(p)}\frac{h_\omega}{2 m_\omega^2} \bar q(-p) \gamma_0 q(p)\,. 
\end{align}
Here, ${\cal V}_4$ is the space-time volume and the four-quark coupling and the Yukawa coupling are related by   
\begin{align}
  \bar\lambda_\omega = \frac{\bar h_\omega^2}{2 m_\omega^2}
  \quad\text{with}\quad \bar h_\omega = \frac{h_\omega}{Z^{1/2}_\omega Z_q} \,. 
	\label{eq:lambdaoho}
\end{align}
In 	\labelcref{eq:lambdaoho}, $\bar \lambda_\omega$ is the RG-invariant coupling \labelcref{eq:lambdanb-lambdaomega} in the density channel. Evidently, $\bar h_\omega$ is the RG-invariant vertex coupling between the density mode $\omega_0$ and the quarks derived from the vertex dressing $\lambda_\omega$, while $m_\omega$ is the mass scale of the density mode $\omega_0$. 

We emphasise that $\omega_0$ specifically refers to the density mode. The $\omega$ vector meson is a subleading four quark tensor structure in the density channel and the temporal component is not a propagating mode with physical polarisation. This subtlety is the reason for the simple kinetic term in \labelcref{eq:Gomega} without a transversal structure as commonly used for vector mesons, see in particular \cite{Rennecke:2015eba, Jung:2016yxl, Jung:2019nnr} for a treatment within the fRG approach and recent developments. 

If combining \labelcref{eq:Gomega} with the emergent fields in the sigma-pion channel with the respective scalar-pseudoscalar field  $\phi=(\sigma,\boldsymbol{\pi})$ within the same approximation used here, we arrive at the following approximation for the matter part of the effective action of QCD for low momenta, 
\begin{align}\nonumber 
	\Gamma_{\textrm{mat}} =&\, \int_p  Z_q(p)  \bar q(p)\, \Biggl[  i\, \pslash -\gamma_0\bar\mu_q + \frac{h_\phi(\rho)}{Z_q(p)} \tau\cdot  \phi \Biggr]\,q(-p) \\[1ex]\nonumber  
	& + \int_p  Z_\phi(p)\,\phi(p) \, p^2 \, \phi(-p) +\int_x V_\textrm{eff}(\rho) \\[1ex]
	& + \frac12\int_p \, Z_\omega(p)\omega_0(-p) \left[ p^2 + m_\omega^2 \right]\,\omega_0(p)  + \cdots\,,
\label{eq:Gmatter}
\end{align}
where $\rho=\tfrac{1}{2} (\sigma^2 +\boldsymbol{\pi}^2)$ and $\tau\cdot \phi= \tfrac{1}{2} ( \sigma + \boldsymbol{\sigma} \boldsymbol{\pi})$ with the scalar-pseudoscalar field \mbox{$\phi=(\sigma,\boldsymbol{\pi})$}. The dots stand for higher order terms, and the shifted chemical potential $\bar\mu_q$ is given by  
\begin{align} 
	\bar \mu_q =\mu_q+\frac{h_\omega}{2 Z_q} \omega_0\,.
\label{eq:barmuq}
\end{align}
 If \labelcref{eq:Gmatter} is evaluated on the equations of motion for $\phi$ and $\omega_0$, the effective action $\Gamma_\textrm{mat}$ reduces to the kinetic term of the quarks and four-quark scattering terms in the scalar-pseudoscalar and vector channels. The coupling $\bar \lambda_\omega$ of the latter is given by \labelcref{eq:lambdaoho}, the coupling $\bar \lambda_\phi$ of the former by 
\begin{align}
  \bar \lambda_\phi  =  \frac{\bar h_\phi^2}{2 m_\phi^2}
  \quad \text{with}\quad \bar \lambda_\phi = \frac{\lambda_\phi}{Z^2_q },\quad \bar h_\phi= \frac{h_\phi}{Z_\phi^{1/2} Z_q }\,, 
	\label{eq:lambdaphihphi}
\end{align}
see e.g.~\cite{Mitter:2014wpa, Braun:2014ata, Rennecke:2015eba, Cyrol:2017ewj, Fu:2019hdw, Fukushima:2021ctq}. In analogy to \labelcref{eq:lambdaoho}, the RG-invariant coupling $\bar h_\phi$ is derived from the vertex dressing $h_\phi(\rho)$. The dynamical $\sigma$ and $\boldsymbol{\pi}$ fields come with a full effective potential $V_\textrm{eff}(\rho)$ that includes all order self-scatterings of these fields.

In turn, it has been shown in \cite{Fukushima:2021ctq} that no such potential is generated for sufficiently small $\omega_0$ due to the property that is often called the ``Silver Blaze'' after Sherlock~Holmes (see \cite{Cohen:2003kd}) for sufficiently small $\omega_0$ and chemical potentials. This connection is apparent in the first line of \labelcref{eq:Gmatter}, where $\omega_0$ enters as a shift of the quark chemical potential, see \labelcref{eq:barmuq}.  Hence, an $\omega_0$-derivative of $\Gamma_\textrm{mat}$ can be written in terms of a $\bar \mu_q$-derivative and the $\omega_0$-derivative of the kinetic term of the $\omega_0$. This leads us to the following equation of motion for a constant $\omega_0$, 
\begin{align}
	\left.\frac{\partial \Gamma_{\textrm{mat}}}{\partial \bar \omega_0}\right|_{\omega_0=\omega_0^\textrm{EoM}} =  \frac{\bar h_\omega}{2} \frac{\partial \Gamma_\textrm{mat}}{\partial \mu_q}+ m_\omega^2 Z^{1/2}_\omega \omega^\textrm{EoM}_0 =0\,,  
	\label{eq:EoMbaromega} 
\end{align}
where we have used that the $\bar \mu_q$-derivative is simply a $\mu_q$-derivative. This makes apparent, that 
the first term on the right hand side is nothing but (minus) the quark density \labelcref{eq:IntDensity}, multiplied by $h_\omega/2Z_q$. Hence, the constant solution of the equation of motion is
\begin{align} 
Z^{1/2}_{\omega}	\omega_0^\textrm{\tiny{EoM}} =\frac{\bar h_\omega}{2 m_\omega^2} n_q\,.
	\label{eq:omega0EoM}
\end{align}
On the equations of motion, \labelcref{eq:omega0EoM} vanishes for chemical potentials smaller than the onset chemical potential $\mu_q^{\text{(on)}}$. To make the lack of $\omega_0$-dependences below onset more apparent, we rewrite the effective action in terms of the shifted and renormalisation group invariant field $\bar\omega_0$ with 
\begin{align}
	\bar\omega_0 = Z_\omega^{1/2} \omega_0 +\frac{2}{\bar h_\omega}\mu_q\,, 
	\label{eq:DefofObar}
\end{align}
where both terms are separately RG-invariant. With \labelcref{eq:DefofObar} we readily obtain
\begin{align}
	\Gamma_\textrm{mat}[Z_\omega^{1/2} \omega_0;\mu_q] = \Gamma_\textrm{mat}[\bar \omega_0;0] -  \frac{2 m_\omega^2}{\bar h_\omega}   \bar\omega_0 \mu_q + \frac{ 2 m_\omega^2}{\bar h_\omega^2}\mu_q^2 \,,
	\label{eq:Gmatterbaromega}
\end{align}
where we have only kept the $\bar\omega_0$- and $\mu_q$-arguments of $\Gamma_\textrm{mat}$ explicit. In $\Gamma_\textrm{mat}[\bar \omega_0;0]$, the field $\bar \omega_0$ has taken the role of the chemical potential and 
hence there is no $\bar \omega_0$-dependence for 
\begin{align}
|\bar \omega_0|\leq \mu_q^{\text{(on)}}\,.
\label{eq:onsetOmega}
\end{align}
This concludes our proof of the absence of an effective potential for $\bar\omega_0$ with \labelcref{eq:onsetOmega}. 

On the solution \labelcref{eq:EoMbaromega} of the EoM, the $\mu_b$-derivative of $\Gamma_\textrm{mat}$ only hits the explicit $\mu_b$-dependence on the right hand side of \labelcref{eq:Gmatterbaromega}, and we arrive at 
\begin{align}
	n_q=\frac{2 m_\omega^2}{\bar h_\omega}   \bar\omega_0^\textrm{\tiny{EoM}} -\frac{4 m_\omega^2}{\bar h_\omega^2}\mu_q = \frac{2 m_\omega^2}{\bar h_\omega}  Z_\omega^{1/2} \omega_0^\textrm{\tiny{EoM}} \,, 
	\label{eq:nqbaromega} 
\end{align}
which vanishes for $\omega^\textrm{EoM}_0=0$. For $\muq> \mu_q^{\text{(on)}}$, the baryon density is non-vanishing. The onset chemical potential is determined by the distance of the first singularity of $M_q(p)$ in the complex frequency plane from the Euclidean frequency axis for vanishing spatial momentum. Since the quark does not define an asymptotic state this singularity may in principle be complex. We evaluate the location of this singularity in \Cref{app:Pade}. Finally, we can feed back the solution \labelcref{eq:nqbaromega} of the EoM into the shifted chemical potential defined in \labelcref{eq:barmuq}. This leads us to 
\begin{align}
	\bar\mu_q^\textrm{\tiny{EoM}}= \mu_q + \frac12 \bar \lambda_\omega\, n_q \,.
\label{eq:barmuEoM}
\end{align}
On the equations of motion the shifted chemical potential only depends on the RG-invariant four-quark coupling $\bar \lambda_\omega$ and the quark number density.

\subsection{The nuclear binding energy from emergent nucleons}

If we reformulate the low-energy effective theory in terms of nucleons, the intricacies due to the complex nature of singularity in the quark propagator are softened by the fact that the nucleon defines an asymptotic state. The respective propagator shows a real pole at the nucleon pole mass, and the computation of the onset chemical potential, the corresponding saturation density and the binding energy as well as further observables is facilitated. Then, the matter part of the effective action is given by 
\begin{align}
	\Gamma_\textrm{mat} \to \Gamma_\textrm{mat} + \Gamma_\textrm{nuc} \,,
\end{align}
with the nucleon term 
\begin{align} \nonumber
&	\Gamma_\textrm{nuc} \\[1ex]
& \approx \int_p  \,Z_b(p) \, \bar{b}(-p)
	\Biggl[ i\, \pslash + \gamma_0 \left( \mu_b+\frac{h_{b\omega}}{2 Z_n} \omega_0 \right)+ m_b \Biggr] b(p) \,,
\label{eq:preGnuc1}
\end{align}
where $b$ is a baryon field, representing nucleons here, and we have restricted ourselves a constant nucleon mass function $M_b(p) =M_b$. Note that in this approximation the Euclidean mass $M_b$ is identical with the pole mass $m_b$ of the nucleon, i.e., $M_b=m_b$. Moreover, we shall also use a constant wave function renormalisation $Z_b\equiv 1$. 

It is left to determine the $b-\omega_0$ coupling $\bar h_{b\omega}$, which is fixed uniquely by the onset condition (that is the Silver Blaze property below the onset), 
\begin{align}
	 \bar h_{b\omega} = 3  \bar h_{\omega}\,, \qquad 
	 \label{eq:hbo}
\end{align}
as has been shown in \cite{Fukushima:2021ctq}.
Below the onset all correlation functions are given by 
\begin{align}
	\Gamma^{(n)}_{\Phi_{i_1}\cdots \Phi_{i_n}}({p_0}_1,...,{p_0}_{n}; \mu_q) = \Gamma^{(n)}_{\Phi_{i_1}\cdots \Phi_{i_n}}({\bar p}_0{}^{\ }_1,...,{{\bar p_0}}{}^{\ }_{n}; 0) \,,
\end{align}
with 
\begin{align}
	{\bar p_0}{}^{\ }_{\Phi_i} = {\bar p_0}{}^{\ }_{\Phi_i}+i\, \alpha_{\Phi_i}\mu_q\,. 
	\label{eq:baromega}
\end{align}
Here, $3 \alpha_{\Phi_i}$ is the baryon number of the respective field. For clarity, we briefly collect the $\alpha$'s relevant for the present work, quark/anti-quark: $\alpha_{q}=\pm 1$, mesons ($\phi,\omega_0$): $\alpha_\textrm{mes}=0$, and nucleon/anti-nucleon:  $\alpha_{b}=\pm 3$. In the presence of the density mode, the quark chemical potential is shifted by $\omega_0$, see \labelcref{eq:barmuq}. Then, \labelcref{eq:baromega} depends on $\bar \mu_q$ instead of $\mu_q$. Specifically, for the baryon field we obtain 
\begin{align}\label{eq:muBar}
	\bar\mu_b = 3\left( \mu_q +\frac{\bar h_\omega}{2}  Z_\omega^{1/2}\omega_0 \right)\,, 
\end{align}
leading to \labelcref{eq:hbo}. In summary this amounts to 
\begin{align}
	\left. \frac{\partial \Gamma_\text{mat}}{\partial \bar \omega_0}\right|_{ \omega_0= \omega^\textrm{\tiny{EoM}} _0} 
  =  \frac{3\bar h_\omega}{2}\frac{\partial \Gamma_\text{mat}}{\partial{\mu}_b} + m_\omega^2  Z^{1/2}_\omega \omega_0^{\text{EoM}} =0\,, 
\end{align}
see also \labelcref{eq:EoMbaromega}. The $\mu_b$-derivative of the effective action is (minus) the nucleon density, 
\begin{align}
	n_b = -\frac{\partial \Gamma_\text{mat}}{\partial{\mu}_b} \,,
\end{align}
and we are led to 
\begin{align}
Z_\omega^{1/2}	\omega^\textrm{EoM}_0=\frac{3h_\omega}{2m_\omega^2}n_b
  \quad \rightarrow \quad   
	\bar{\mu}_b=\mu_b+2 \lambda_{n_b n_b} n_b\,, 
\label{eq:EoMomega}
\end{align}
where we have used 	\labelcref{eq:lambdanb-lambdaomega}. This is the nucleon analogue of \labelcref{eq:barmuEoM}. 
As we emphasise there, $\bar \mu_{b}$ does not depend on dynamical hadronisation parameters $m_\omega, h_\omega$ on the EoM{}. Instead, it only depends on the nucleon density as well as the fundamental four-nucleon scattering coupling $\lambda_{n_b n_b}$ in the density-channel. 

We assume that off-shell fluctuations of the nucleon play no relevant role to describe the onset of nuclear matter. Moreover, the quark part of the density is subleading nature, and we drop it in the following. Then, the density fluctuations are given by
\begin{align}\nonumber 
	\Gamma_{\bar{\mu}_b}=&\frac{1}{12\pi^2}  \Biggl[{\bar{\mu}_b}
	\left( 5 m_b^2-2 {\bar{\mu}_b}^2\right) \sqrt{{\bar{\mu}_b}^2-m_b^2}\\[2ex]
	& \hspace{-.5cm}+\frac{3}{2} m_b^4 \log \left( \frac{\bar\mu_b
   - \sqrt{\bar\mu_b^2  - m_b^2}}{\bar\mu_b + \sqrt{\bar\mu_b^2 - m_b^2}}\right)\Biggr]\theta({\bar{\mu}_b}^2-m_b^2)\,.
	\label{eq:Gmu}
\end{align}
Differentiation with respect to the nucleon chemical potential yields a self-consistency equation for the nuclear density
\begin{align}
	n_{b}^2 =&\,\frac{ 4}{(3 \pi ^2)^2}
	\left({\bar{\mu}_b}^2-m_b^2\right)^{3}
	\label{eq:NucleonDensity}
\end{align}
with the density-dependent $\bar \mu_{b}$ in \labelcref{eq:EoMomega}. At the first order phase transition into the nuclear liquid, the effective nucleon mass is expected to jump to a screened, smaller value as a consequence of a jump of the chiral condensate $\sigma$ to a smaller but still finite value. We arrive at a self-consistency equation for the onset value of the chemical potential $\mu^\ast_b$ or the nucleon mass $m^\ast_b$ at saturation density, 
\begin{align}
	\mu^\ast_b=\sqrt{\bigg( \frac{3\pi^2 n^{\text{(sat)}}_b}{2}\bigg)^{\frac{2}{3}} + \left(m^\ast_b\right)^2}-2 \lambda_{n_b n_b}^{\text{(sat)}} \, n_b^\text{(sat)}\,.
\label{eq:onsetmu}
\end{align}
For the vector coupling $\lambda_{n_b n_b}^{\text{(sat)}}$ at saturation density we use the minimum of $V_b(\vec p{\,}^2)$, depicted in \Cref{fig:VectorPotentialMom}, which leads us to 
\begin{align}
\lambda_{n_b n_b}^{\text{(sat)}}\approx -\SI{120}{GeV^{-2}}\,, 
\label{eq:Minimumlambadnb}
\end{align}
related to the minimum in position space used to determine the saturation density \labelcref{eq:density_lattice}. Moreover, with \labelcref{eq:binding-energy} we get \mbox{$\mu^\ast_b = \SI{918}{\MeV}$}, which was obtained by using the vacuum nucleon mass \mbox{$m_b = \SI{939}{\MeV}$}; see~\Cref{app:BaryonPoleMass}. With this input the self-consistency equation can be solved numerically for $m_b^\ast$ as a function of $\mu^\ast_b$, which leads us to 
\begin{align}
m_b^\ast\approx 469\,\textrm{MeV}\,, 
\label{eq:Mbliquid}
\end{align}
which shows $50\%$ reduction of the in-medium nucleon mass as compared to the vacuum mass.
This estimate is consistent with the conventional calculations within
the Dirac-Brueckner-Hartree-Fock theory; see \cite{vanDalen:2005ns}
for the relativistic/non-relativistic definitions of the effective
masses as functions of the Fermi momentum, \cite{Baldo:2014yja} for
astrophysical implications, \cite{Shang:2020kfc} for
finite-temperature extension, and references therein.

\section{Conclusions}
\label{sec:Conclusions}

We have set-up a first principles approach for computing properties of the nuclear liquid-gas transition in QCD\@. This approach is based on the computation of the potentials of density correlations in QCD in terms of correlation functions in first principles QCD\@. Specifically we have related the potential of density-density correlations to the four-quark scattering coupling $\bar \lambda_\omega$ in the vector channel. From this potential we deduced estimates for the saturation density $\nsat$, see \labelcref{eq:density_lattice}, and the binding energy $\epsilon_b$, see \labelcref{eq:binding-energy}, in \Cref{sec:BindingEnergyDDPot}, including a systematic error estimate of 25\%. 

In \Cref{sec:BindingEnergyLEFT} we have set-up a QCD-assisted low energy effective action, using results in \cite{Fukushima:2021ctq}, whose vertex dressings and kinetic terms are directly related to correlation functions in first principles QCD\@. The key ingredient is again given by $\bar \lambda_\omega$. As a self-consistency check we have computed the nucleon mass in the nuclear liquid, which turned out to be about $50\%$ of the nucleon mass in the vacuum, see \labelcref{eq:Mbliquid}. 

The qualitative results in the present work are very encouraging, and
their systematic improvement towards quantitative precision is subject
of current work. Updates of the onset baryon chemical potential, the saturation
density, the nucleon mass in the liquid as well as other observables
await further investigations in the near future.

\section*{Acknowledgements}

We thank Fabian Rennecke for discussions. JH and JMP thank the University of Tokyo and the Yukawa Institute for Theoretical Physics Kyoto for hospitality, where this work was finalised. This work is funded by the Deutsche Forschungsgemeinschaft (DFG, German Research Foundation) under Germany’s Excellence Strategy EXC 2181/1 - 390900948 (the Heidelberg STRUCTURES Excellence Cluster) and the Collaborative  Research Centre SFB 1225 - 273811115 (ISOQUANT). NW acknowledges support by the Deutsche Forschungsgemeinschaft (DFG, German Research Foundation) – Project number 315477589 – TRR 211 and by the State of Hesse within the Research Cluster ELEMENTS (Project ID 500/10.006). KF is supported by JSPS KAKENHI Grant Nos.\ 22H01216 and 22H05118.


\appendix

\section{Fit for the four-quark scattering coupling $\bar \lambda_{\omega}(p)$ in the vector channel}
\label{app:FitVb}

\begin{table*}[]
	\renewcommand{\arraystretch}{1.5}
	\begin{tabular}{|c|c|c|c|c|c|c|c|}
		\hline
		& $a_0$ & $a_1\left[\textrm{MeV}^{-2}\right]$ & $a_3\left[\textrm{MeV}^{-2}\right]$ & $b_0\left[\textrm{MeV}^{2}\right]$ & $b_1$ &$b_2\left[\textrm{MeV}^{-2}\right]$ & $b_3\left[\textrm{MeV}^{-4}\right]$\\ \hline
		$\bar\lambda_\omega$ & $1$& $-1.01\cdot10^{-4}$ & $-7.45\cdot 10^{-12}$ & $6.51\cdot10^{5}$ & $0.692$ &$3.19\cdot10^{-6}$ & $9.18\cdot10^{-12}$\\ \hline
		$\bar\lambda_{\omega,\textrm{low}}$&$1$&$-1.76\cdot10^{-4}$&$-5.98\cdot10^{-12}$&$9.34\cdot10^{5}$&$0.975$&$3.66\cdot10^{-6}$&$1.63\cdot10^{-11}$\\ \hline
		$\bar\lambda_{\omega,\textrm{up}}$&$1$&$-0.88\cdot10^{-4}$&$-5.87\cdot10^{-12}$&$5.85\cdot10^{5}$&$0.682$&$3.45\cdot10^{-6}$&$7.56\cdot10^{-12}$\\ \hline
		\end{tabular}
	\caption{Fit parameters for the vector coupling $\bar\lambda_\omega(p)$ as prescribed by \labelcref{eq:lambda_shape}. Its nucleon analogue $\lambda_{n_b n_b}(p)=9/4 \bar\lambda_\omega$, see \labelcref{eq:lambdanb-lambdaomega}, is shown in \Cref{fig:VectorPotentialMom}. The upper and lower bounds for the vector coupling are inherited from the original QCD calculation \cite{Cyrol:2017ewj}.}
	\label{tab:FitPar}
\end{table*}

In this Appendix we provide a quantitative Padé fit of the QCD data of the density channel dressing,
\begin{align}\label{eq:lambda_shape}
	\bar \lambda_{\omega}(p)=\frac{a_0+a_1p^2+a_2p^4}{b_0+b_1p^2+b_2p^4+b_3p^6} \,.
\end{align}
from which the potential is obtained via~\labelcref{eq:VbFourier,eq:lambdanb-lambdaomega}. The fit is obtained from a least $\chi^2$ optimisation. The fit parameters $a_i$ and $b_i$ are listed in \Cref{tab:FitPar}.

\section{Baryon pole mass}\label{app:BaryonPoleMass}

In this Appendix we infer bounds on the nucleon pole mass $m_b$ from the position of the first singularity in the complex momentum plane of the quark propagator. The present analysis is based on the QCD approach with emergent composites as set-up in \cite{Fukushima:2021ctq}, including the density mode $\omega_0$, see \Cref{sec:BindingEnergyLEFT}. 

We start by noting that the quark density operator is given by the expectation value of the quark number operator, and hence obeys
\begin{align} 
	n_q(\mu) \propto \int \frac{d^4 p}{(2\pi)^4}\,\tr \, \gamma_0\, G_q(p) \,.
\label{eq:density-op}
\end{align}
The frequency integral in \labelcref{eq:density-op} vanishes in the vacuum as $\tr \, \gamma_0 G_q$ is an odd function in frequency space. For small chemical potentials one can always absorb $\bar\mu_q$ in the frequency integral. However, if $\omega + i \bar \mu_q$ sweeps over the first singularity in the complex plane, the density starts rising. 

The quark propagator has two branches, $G^-$ below the onset chemical potential and $G^+$ above the onset chemical potential, with $\bar \mu_q \gtrless  \mu_q^\textrm{(on)}$. It reads 
\begin{align} \label{eq:quark-prop+-}
	G^\pm_q(p,\mu_q) = \frac{1}{Z^\pm_q(\bar p)} \frac{-i \slashed{\bar p}  + M^\pm_q(\bar p)}{\bar p^2 + M^\pm_q{}^2(\bar p)} \,, 
\end{align}
with 
\begin{align}
	\bar p_\mu= p_\mu + \imag \delta_{\mu 0} \bar\mu_q\,,
\end{align}
and $\bar \mu_q$ in \labelcref{eq:barmuq}. While $Z^+_q, M^+_q$ have a genuine $\mu_q,\omega_0$-dependence, $Z^-_q, M^-_q$ are that in the vacuum due to Silver blaze, and hence 
\begin{align}
	G^-(p,\mu)=G^-(\bar p,0)\,.
\end{align}
Both propagators carry singularities in the complex frequency plane with the positions $\omega_\textrm{sing}$ and we define 
\begin{align}
	m^\pm_\textrm{sing}= \min \big| \textrm{Re} (\omega^\pm_\textrm{sing} ) \big|\,,  
	\label{eq:msing}
\end{align}
with 
\begin{align} 
 m^+_\textrm{sing}  \leq \mu_q^\textrm{(on)}\leq 	m^-_\textrm{sing}\,. 
 \label{eq:muonmpm}
\end{align}
The onset-chemical potential $\mu_q^\textrm{(on)}$ is nothing but the combination of the baryon pole mass $m_{b}$ and the binding energy $\epsilon_{b}$, 
\begin{align} 
 \mu_q^\textrm{(on)} = \frac{1}{3} \big( m_{b} - \epsilon_{b} \big)\,, 
\end{align}
and in conclusion, $3 m_\text{sing}^\pm+\epsilon_{b}$ are lower and upper estimates of the nucleon pole mass, 
\begin{align} 
	m^+ _{\textrm{sing}}   <  \frac{1}{3} \big( m_{b} - \epsilon_{b} \big) < m^- _{\textrm{sing}} \,.
\label{eq:Estimatemuonset}
\end{align}
\begin{figure*}[t]
	\centering
	\subfloat{\includegraphics[width=.3\textwidth]{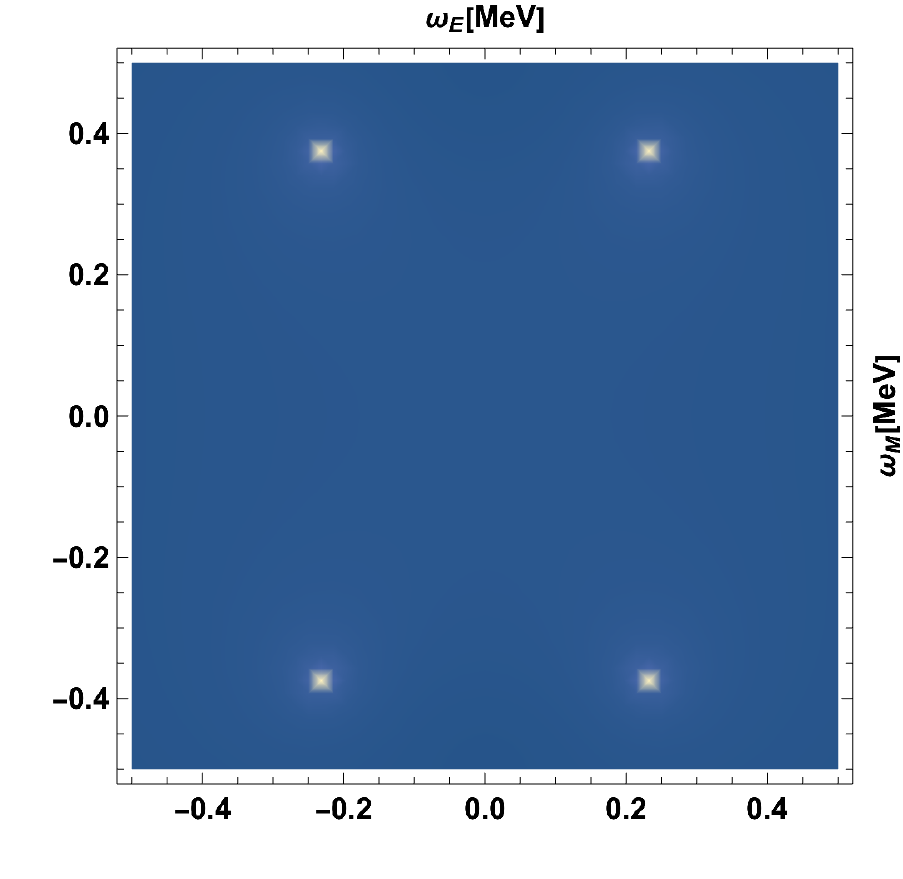}}
	\subfloat{
		\includegraphics[width=.3\textwidth]{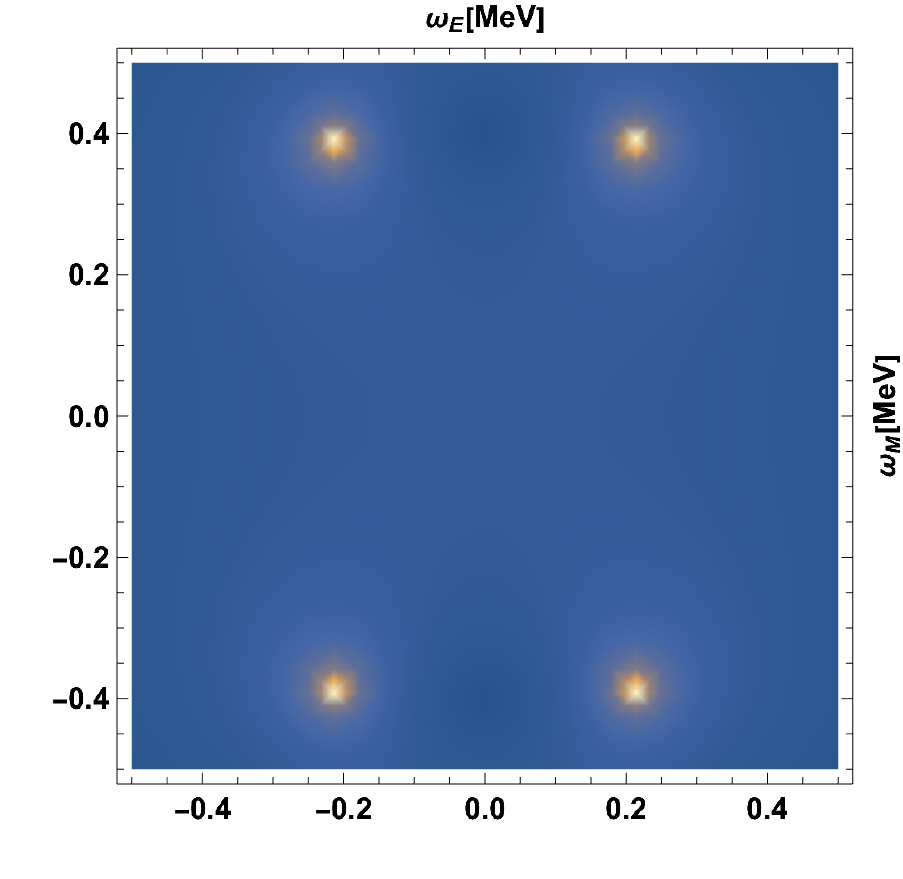}}
	\subfloat{
		\includegraphics[width=.3\textwidth]{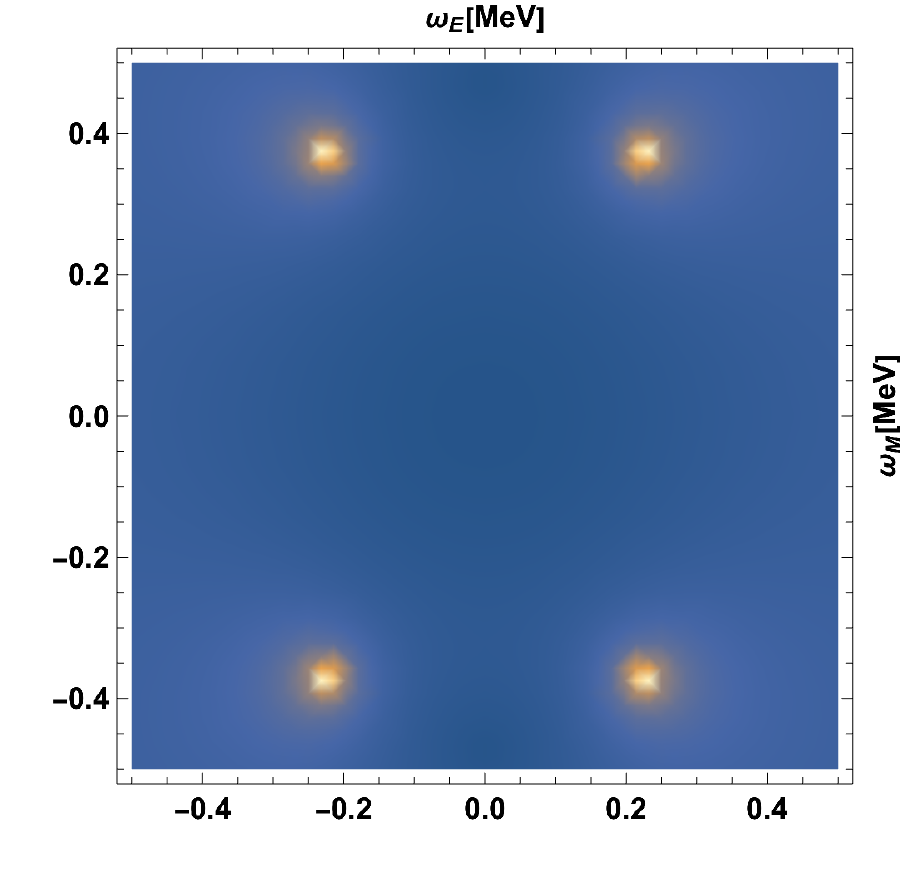}}	
	\caption{Pad\'{e} continuation of $g_s$~\labelcref{eq:univ-part} (left), $g_s/Z$ (middle) and $M g_s$ (right) in the complex plane. The distance of the complex conjugate poles to the Euclidean axis yields an upper estimate of the nucleon pole mass with~\labelcref{eq:Estimatemuonset}, while the deviation of the pole position in the displayed reconstructions yields an error estimate.}
	\label{fig:pade-quark}
\end{figure*}

In \Cref{app:Pade}, we use numerical results for the Euclidean vacuum quark propagator in two-flavor QCD from~\cite{Cyrol:2017ewj} to determine the position of the first complex singularity $m^-_\text{sing}$ of the quark propagator in the below-onset-branch. Due to lack of data in the upper branch, we can only provide an upper estimate for the nucleon mass with~\labelcref{eq:Estimatemuonset}. 

To arrive at this estimate, we need to relate our result for to the two-flavor quark pole mass \labelcref{eq:msing2017} to the corresponding 2+1-flavor QCD result used in~\labelcref{eq:Estimatemuonset}. We do this by assuming that the ratio of the pole masses is the same as the ratio of the constituent light quark masses $M_l$,
\begin{align} \label{eq:const-pole-ratio}
	\frac{m_\text{sing}^\text{(2)}}{M_l^\text{(2)}} = \frac{m_\text{sing}^\text{(2+1)}}{M^\text{(2+1)}} \,,
\end{align}
indicating the number of flavors of the results in the respective superscripts. The reconstruction of the two-flavor pole mass based on the Euclidean two-flavor data in \cite{Cyrol:2017ewj} is detailed in \Cref{app:Pade}, its value being $\SI{377\pm 9}{\MeV}$, see  \labelcref{eq:cc-pole}. The respective two-flavor constituent quark mass from~\cite{Cyrol:2017ewj} is given by   $m_\text{const}^\text{(2)} = \SI{418\pm42}{\MeV}$, and the 2+1-flavor constituent mass from ~\cite{Gao:2021wun} is given by \mbox{$ M_l^\text{(2+1)} = \SI{350\pm35}{\MeV} $}. The assigned 10\% error is the conservative estimate for the systematic errors in \cite{Cyrol:2017ewj,Gao:2021wun}. Using~\labelcref{eq:const-pole-ratio} in~\labelcref{eq:Estimatemuonset} yields
\begin{align} \label{eq:nucleon-mass_2f}
	m_b < 3 \, \text{Re} \Bigg[ m_\text{sing}^{(2)} \frac{M_l^\text{(2+1)}}{M_l^\text{(2)}} \Bigg] + \epsilon_b \,,
\end{align}
which, inserting our Pad\'e reconstruction result~\labelcref{eq:cc-pole} as well as our estimate for the binding energy~\labelcref{eq:binding-energy}, leads us to our estimate for the nucleon mass
\begin{align} \label{eq:nucleon-mass_estimate}
	m_b < \SI{967\pm97}{\MeV} \,,
\end{align}
which is in good agreement with the experimental value of $m_b^\text{exp} = \SI{938}{\mega\eV}$.

\section{Analytic continuation of the quark propagator} \label{app:Pade}

In this appendix, we extract the leading singularity of the quark propagator by fitting the corresponding Euclidean data with a rational function (Pad\'e approximation). In general, if being short of numerical data for correlation functions for timelike momenta, one employs reconstruction methods for deducing the correlation functions for timelike momenta from the numerical data of the Euclidean one. This generically ill-conditioned problem has been studied for QCD correlation functions in~\cite{Haas:2013hpa,Ilgenfritz:2017kkp, Cyrol:2018xeq, Dudal:2019gvn, Li:2019hyv, Hayashi:2021nnj,Horak:2021syv, Horak:2023xfb} and for the particular case of the quark in~\cite{Karsch:2009tp, Mueller:2010ah, Qin:2013ufa, Fischer:2017kbq}, and is sustained by respective direct computations of timelike correlation functions within functional methods in \cite{Kamikado:2013sia, Fischer:2020xnb, Pawlowski:2015mia, Pawlowski:2017gxj, Horak:2020eng, Horak:2021pfr, Horak:2022myj, Horak:2022aza}. 

Typically, reconstruction methods suffer from considerable systematic errors for secondary or even higher order structures in the complex plane. This relates to the required exponential accuracy for the resolution of sub-leading decays for spacelike Euclidean momenta. In turn, the location (while not the shape) of the first complex structure (smallest distance to the Euclidean frequency axis) can be obtained from most reconstruction methods with a small systematic error. 

Hence, to extract the baryon mass from the complex structure of the quark propagator, we use a simple Pad\'{e} approximation for the scalar part of the quark propagator. Note, that Pad\'{e} approximants face even more serious conceptual problems as a reconstruction method than some other approaches. In particular they cannot describe cuts, which typically are represented by accumulations of poles. However, these problems do not affect the location of the \textit{first} singularity in the complex plane, which is the only information of relevance for the present task of deducing an upper bound on the nucleon pole mass. 

The leading singularity of the quark propagator in the complex $p^2$-plane can be interpreted as a result of the quasiparticle nature of the quark. Hence, the position of the singularity should be linked to the quarks mass scale, encoded in the mass function $M_q$. In consequence, this singularity should be carried by the universal part of the quark propagator, 
\begin{align} 
	g(p)=\frac{1}{p^2+M_q^2(p)} \,.
\label{eq:univ-part}
\end{align}
The universal part $g$ only depends on the mass function $M_q(p)$. The momentum (and RG) scaling of the field, encoded in $Z_q$, has been removed, whose momentum dependence evidently should not introduce poles. 

By constructing Pad\'{e} approximants for the universal part $g$, we extract the leading singularity of the quark propagator. We complement this by Pad\'{e} approximants for $g_q/Z_q$ and $g_q M_q$ to construct at error estimate for the position of the singularity. Furthermore, this allows to check that wave function $Z_q$ does not introduce additional poles. 

In \Cref{fig:pade-quark}, we present the Pad\'{e} approximants of $g_q$, $g_q/Z_q$ and $g_q M_q$ in the complex plane. We clearly identify complex poles for all approximants, and extract the pole position $\omega_\textrm{sing}$ with 
\begin{align}
	\omega_\text{sing} = 377(9) \pm \imag\, 22(1) \, \text{MeV} \,, 
 \label{eq:cc-pole}
 \end{align}
which leads to $m_\textrm{sing}$ in \labelcref{eq:msing} with 
\begin{align}
		m_\text{sing} = \SI{377\pm 9}{MeV}\,.
		\label{eq:msing2017}
\end{align}
The corresponding constituent quark mass of the data \cite{Cyrol:2017ewj} is $m_\text{const} = M_q(0) = \SI{418}{\MeV}$. The Padé analysis only provides a reliable information about the location of the first singularities in the complex plane as already discussed. With rational approximants these singularities are always represented by poles. Still, assuming their actual existence as poles, the respective spectral representation of $g_q(p)$ reads 
\begin{align}
	g_q(p) = \frac{R}{p^2+\omega_\textrm{sing}^2} +  \frac{\bar R}{p^2+\bar \omega_\textrm{sing}^2} +  \int_0^\infty \frac{\mathrm{d} \lambda}{\pi} \frac{\lambda\, \tilde \rho_g(\lambda)}{p^2 + \lambda^2} \,,
\label{eq:gqcc}
\end{align}
where $\bar R,\bar \omega_\textrm{sing}$ are the complex conjugates of $R,\omega_\textrm{sing}$ and $\tilde\rho_g(\lambda)$ carries the scattering tail of $g_q(p)$. 

The result of~\labelcref{eq:msing2017} can be compared to the results of~\cite{Horak:2022aza}, where a direct calculation of the vacuum quark spectral function was put forward by using the spectral DSE approach. This approach also encompasses the scenarios with complex conjugate poles such as \labelcref{eq:gqcc}, for a detailed discussion see~\cite{Horak:2022aza}. In the absence of complex conjugate poles, the spectral representation for the universal part \labelcref{eq:univ-part} 
\begin{subequations}\label{eq:spec-rep_univ-prop}
\begin{align} 
	g(p) = \int_0^\infty \frac{\mathrm{d} \lambda}{\pi} \frac{\lambda\, \rho_g(\lambda)}{p^2 + \lambda^2} \,,
	\label{eq:Specg}
\end{align}
with the corresponding spectral function
\begin{align} \label{eq:univ-spec}
	\rho_g(\omega) = 2 \, \text{Im} \, g(-\imag \omega + 0^+) \,.
\end{align}
\end{subequations}
In~\cite{Horak:2022aza}, it was observed that an analytic pole-tail split into distributional and continuous contributions describes the quark spectral function extremely well. Within this split, the universal spectral function can be parametrised as
\begin{align}
	\rho_g(\omega) = \frac{\pi}{m_\text{sing}} R \, \delta(\omega - m_\text{sing}) + \tilde \rho_g(\omega) \,,
 \label{eq:split_univ-spec}
 \end{align}
for $\omega \geq 0$, and $\rho_g(-\omega) = - \rho_g(\omega)$. In \labelcref{eq:split_univ-spec}, $R$ denotes the positive residue of the pole at $m_\text{sing}$. $\tilde \rho_g$ captures continuous scattering contributions. We emphasise that $\rho_g$ is renormali sation group invariant as are both, $R$ and $\tilde \rho_g$, cannot be changed via a reparametrisation and hence carry direct physics information. Indeed, they determine the ratio of the constituent mass $m_\textrm{sign}$ and $m_\textrm{const}$. The universal spectral function obeys the sum rule 
\begin{align}
	R + \int_{m_\textrm{sign}}^\infty \frac{\mathrm{d} \lambda}{\pi} \,\lambda \,\tilde \rho_g(\lambda)=1 \,, 
\label{eq:SumRule}
\end{align}
dictated by the decay behaviour $g(p^2\to \infty) \to 1/p^2$. The sum rule \labelcref{eq:SumRule} also persists in the presence of cc poles with $R\to R+\bar R$. In either case, the value of $R>0$ is not bounded by unity as in the case of the pole contribution of physical spectral functions, as the scattering tail $\tilde \rho_g$ contains negative parts. For example, in the direct spectral computation in~\cite{Horak:2022aza}, the pole position $m^\text{DSE}_\text{sing}$ and residue $R^\text{DSE}$ of the light flavors was determined to be 
\begin{align} \label{eq:pole-res}
	m^\text{DSE}_\text{sing} \approx \SI{485}{\mega\eV} \,,\qquad 
	R^\text{DSE} \approx 2.25 \,.
\end{align}
Note that the residue in~\labelcref{eq:pole-res} differs from those of the scalar and vector component in~\cite{Horak:2022aza} of $1/Z_q$ resp. $M/(m_\text{sing} Z_q)$. 

\begin{figure} 
	\centering
	\includegraphics[width=.95\linewidth]{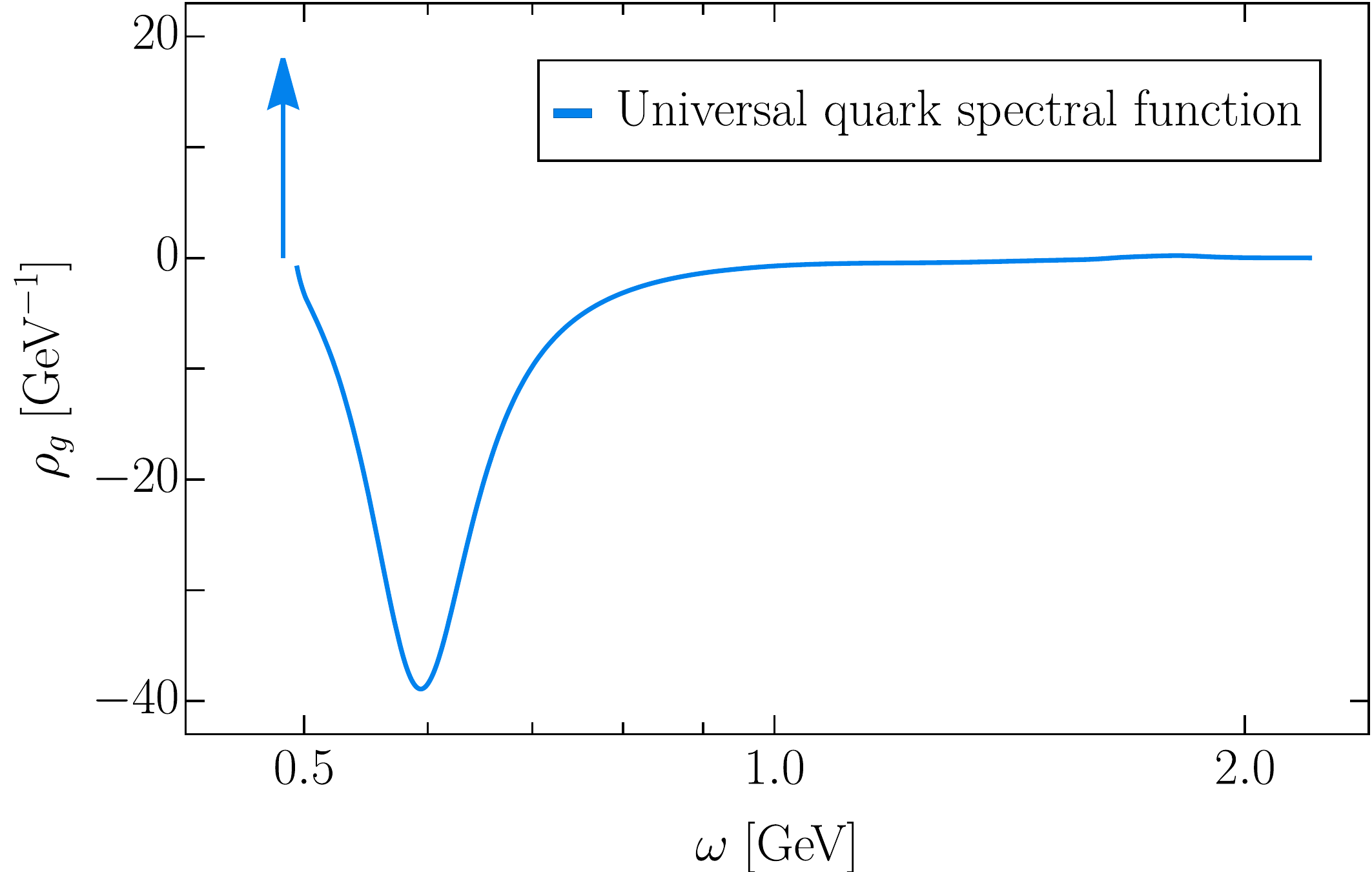}
	\caption{Universal quark spectral function as defined in~\labelcref{eq:spec-rep_univ-prop} for light quark flavors in an isospin symmetric approximation from the spectral DSE results of \cite{Horak:2022aza}. The pole position is $m_\text{sing} \approx \SI{485}{\mega\eV}$. The continuous tail contributes negatively to the mass function of the quark, resulting in a constituent quark mass of $M_l \approx \SI{400}{\mega\eV}$.}
	\label{fig:univ-spec}
\end{figure}
The respective spectral function is depicted in \Cref{fig:univ-spec}. Notably, the pole mass is larger than its corresponding constituent quark mass $M_l^\text{DSE} \approx \SI{400}{\mega\eV}$, in contrast to the result in~\labelcref{eq:msing2017}. The hierarchy of the masses found in~\cite{Horak:2022aza} agrees with that obtained via vertex models widely used in bound state calculations with BSE-DSE systems: the real part of the leading complex singularity in the quark propagator is typically found to be considerably larger than the respective constituent mass~\cite{Eichmann:2016yit, Windisch:2016iud}.  In turn, the Euclidean data that underlie the reconstruction leading to~\labelcref{eq:msing2017} are obtained in a qualitatively better approximation than that used in the direct realtime computation in \cite{Horak:2022aza}. 

Together this hints at a strong  truncation dependence of the relation between quark constituent and pole mass which asks for a refined analysis. Such an analysis is beyond the scope of the present work and we only briefly discuss the underlying dynamics. For the sake of simplicity we restrict ourselves to the scenario with a spectral representation \labelcref{eq:SumRule} and only comment on minor modifications in the general case \labelcref{eq:gqcc} including also cc poles. With \labelcref{eq:univ-part,eq:Specg,eq:split_univ-spec}, the constituent quark mass $m_\text{const}=M_q(0)$ and $m_\text{sing}$ are related by 
\begin{align}
m_\textrm{sing}^2={\cal C} 	m^2_\text{const}\,,
\label{eq:msingmconst}
\end{align}
with 
\begin{align}
{\cal C}=	{R +  m_\textrm{sing}^2 \int_{m_\textrm{sign}}^\infty \frac{\mathrm{d} \lambda}{\pi} \frac{  \tilde \rho_g(\lambda)}{ \lambda}}\,.
	\label{eq:DenomRatio}
\end{align}
If cc poles are present, \labelcref{eq:DenomRatio} is changed by 
\begin{align} 
	R\to \left( \frac{R}{\omega_\textrm{sing}^2}  + \frac{\bar R}{\bar \omega_\textrm{sing}^2}\right)\, m_\textrm{sing}^2\,.
\end{align}
\Cref{eq:msingmconst} admits both, solutions with \mbox{$M_l > m_\text{sing}$} and  $M_l < m_\text{sing} $. In comparison to the sum rule \labelcref{eq:SumRule}, the spectral integral over $\tilde \rho_g$ in \labelcref{eq:msingmconst} carries an additional factor $1/\lambda^2$. Hence, the contributions of larger spectral values are suppressed. This is more clearly seen if rewriting the denominator in \labelcref{eq:msingmconst} with help of the sum rule \labelcref{eq:SumRule}, 
\begin{align} 
	{\cal C}=1+	\int_{m_\textrm{sign}}^\infty \frac{\mathrm{d} \lambda}{\pi}\, \lambda\, \tilde \rho_g(\lambda)\left[ \frac{m_\textrm{sing}^2}{\lambda^2}-1\right] \,. 
\label{eq:wightedthilderho}
\end{align}
The integrand in \labelcref{eq:wightedthilderho} is counting the spectral weight of the scattering tail $\tilde \rho_g$, where the expression in the square bracket is an additional negative weight factor. For \labelcref{eq:gqcc}, the relation  \labelcref{eq:wightedthilderho} has to be augmented with further but subleading terms proportional to $\textrm{Im}[ \omega_\textrm{sing}^2]/m_\textrm{sing}^2$. 

\Cref{eq:wightedthilderho} allows for a qualitative analysis of the required dynamics for the two cases. A simple scenario is provided by the spectral function computed in \cite{Horak:2022aza} and depicted in \Cref{fig:univ-spec}. There, the weight of the scattering tail is concentrated at spectral values larger but close to the onset of the tail at $\lambda=  m_\textrm{sing}$. Just above the onset, we have $\tilde \rho_g(\lambda \gtrsim m_\textrm{sing}) \leq 0$ and hence the integral in \labelcref{eq:wightedthilderho} is positive, leading to ${\cal C}\approx 1.21$ and hence $M_l < m_\text{sing} $. 

In turn, ${\cal C}<1$ is achieved for a scenario where the spectral function $\rho_g$ and hence also $\tilde \rho_g$ resembles more that of a physical particle. For the latter case the scattering tail $\tilde \rho_g >0$ and naturally we have ${\cal C}<1$ with $M_l > m_\text{sing} $. Translated to the present case with the quark spectral function, ${\cal C}<1$ is achieved for $\tilde \rho_g$ with a less pronounced or absent negative peak and more positive weight for larger spectral values. Such a scenario requires more genuine scatterings at large spectral values. The definite answer of whether or not these scatterings take place can only be answered in quantitatively reliable approximations that allow for momentum transfers in the vertices. In particular, within the truncation with constant vertex dressings used in \cite{Horak:2022aza} this specific question cannot be answered. In turn, while the truncation scheme used in \cite{Cyrol:2017ewj,Gao:2021wun} is state-of-the-art, these computations are done in the Euclidean domain and the order of the constituent quark mass $M_l$ and $m_\text{sing} $ is obtained within a reconstruction of the pole position in the complex plane. Still, we consider the latter order as a strong hint for  $m_\text{sing}<M_l$, but leave the decisive analysis to future work.

\bibliographystyle{apsrev4-2}
\bibliography{diquark}

\begin{thebibliography}{50}%
\makeatletter
\providecommand \@ifxundefined [1]{%
 \@ifx{#1\undefined}
}%
\providecommand \@ifnum [1]{%
 \ifnum #1\expandafter \@firstoftwo
 \else \expandafter \@secondoftwo
 \fi
}%
\providecommand \@ifx [1]{%
 \ifx #1\expandafter \@firstoftwo
 \else \expandafter \@secondoftwo
 \fi
}%
\providecommand \natexlab [1]{#1}%
\providecommand \enquote  [1]{``#1''}%
\providecommand \bibnamefont  [1]{#1}%
\providecommand \bibfnamefont [1]{#1}%
\providecommand \citenamefont [1]{#1}%
\providecommand \href@noop [0]{\@secondoftwo}%
\providecommand \href [0]{\begingroup \@sanitize@url \@href}%
\providecommand \@href[1]{\@@startlink{#1}\@@href}%
\providecommand \@@href[1]{\endgroup#1\@@endlink}%
\providecommand \@sanitize@url [0]{\catcode `\\12\catcode `\$12\catcode
  `\&12\catcode `\#12\catcode `\^12\catcode `\_12\catcode `\%12\relax}%
\providecommand \@@startlink[1]{}%
\providecommand \@@endlink[0]{}%
\providecommand \url  [0]{\begingroup\@sanitize@url \@url }%
\providecommand \@url [1]{\endgroup\@href {#1}{\urlprefix }}%
\providecommand \urlprefix  [0]{URL }%
\providecommand \Eprint [0]{\href }%
\providecommand \doibase [0]{https://doi.org/}%
\providecommand \selectlanguage [0]{\@gobble}%
\providecommand \bibinfo  [0]{\@secondoftwo}%
\providecommand \bibfield  [0]{\@secondoftwo}%
\providecommand \translation [1]{[#1]}%
\providecommand \BibitemOpen [0]{}%
\providecommand \bibitemStop [0]{}%
\providecommand \bibitemNoStop [0]{.\EOS\space}%
\providecommand \EOS [0]{\spacefactor3000\relax}%
\providecommand \BibitemShut  [1]{\csname bibitem#1\endcsname}%
\let\auto@bib@innerbib\@empty
\bibitem [{\citenamefont {Bertsch}\ and\ \citenamefont
  {Siemens}(1983)}]{Bertsch:1983uv}%
  \BibitemOpen
  \bibfield  {author} {\bibinfo {author} {\bibfnamefont {G.}~\bibnamefont
  {Bertsch}}\ and\ \bibinfo {author} {\bibfnamefont {P.~j.}\ \bibnamefont
  {Siemens}},\ }\bibfield  {title} {\bibinfo {title} {{NUCLEAR
  FRAGMENTATION}},\ }\href {https://doi.org/10.1016/0370-2693(83)90004-7}
  {\bibfield  {journal} {\bibinfo  {journal} {Phys. Lett. B}\ }\textbf
  {\bibinfo {volume} {126}},\ \bibinfo {pages} {9} (\bibinfo {year}
  {1983})}\BibitemShut {NoStop}%
\bibitem [{\citenamefont {Chomaz}(2002)}]{Chomaz:2002xxf}%
  \BibitemOpen
  \bibfield  {author} {\bibinfo {author} {\bibfnamefont {P.}~\bibnamefont
  {Chomaz}},\ }\bibfield  {title} {\bibinfo {title} {{The nuclear liquid gas
  phase transition and phase coexistence}},\ }\href
  {https://doi.org/10.1063/1.1469927} {\bibfield  {journal} {\bibinfo
  {journal} {AIP Conf. Proc.}\ }\textbf {\bibinfo {volume} {610}},\ \bibinfo
  {pages} {167} (\bibinfo {year} {2002})},\ \Eprint
  {https://arxiv.org/abs/nucl-ex/0410024} {arXiv:nucl-ex/0410024} \BibitemShut
  {NoStop}%
\bibitem [{\citenamefont {Eichmann}\ \emph {et~al.}(2016)\citenamefont
  {Eichmann}, \citenamefont {Sanchis-Alepuz}, \citenamefont {Williams},
  \citenamefont {Alkofer},\ and\ \citenamefont {Fischer}}]{Eichmann:2016yit}%
  \BibitemOpen
  \bibfield  {author} {\bibinfo {author} {\bibfnamefont {G.}~\bibnamefont
  {Eichmann}}, \bibinfo {author} {\bibfnamefont {H.}~\bibnamefont
  {Sanchis-Alepuz}}, \bibinfo {author} {\bibfnamefont {R.}~\bibnamefont
  {Williams}}, \bibinfo {author} {\bibfnamefont {R.}~\bibnamefont {Alkofer}},\
  and\ \bibinfo {author} {\bibfnamefont {C.~S.}\ \bibnamefont {Fischer}},\
  }\bibfield  {title} {\bibinfo {title} {{Baryons as relativistic three-quark
  bound states}},\ }\href {https://doi.org/10.1016/j.ppnp.2016.07.001}
  {\bibfield  {journal} {\bibinfo  {journal} {Prog. Part. Nucl. Phys.}\
  }\textbf {\bibinfo {volume} {91}},\ \bibinfo {pages} {1} (\bibinfo {year}
  {2016})},\ \Eprint {https://arxiv.org/abs/1606.09602} {arXiv:1606.09602
  [hep-ph]} \BibitemShut {NoStop}%
\bibitem [{\citenamefont {Brambilla}\ \emph {et~al.}(2014)\citenamefont
  {Brambilla} \emph {et~al.}}]{Brambilla:2014jmp}%
  \BibitemOpen
  \bibfield  {author} {\bibinfo {author} {\bibfnamefont {N.}~\bibnamefont
  {Brambilla}} \emph {et~al.},\ }\bibfield  {title} {\bibinfo {title} {{QCD and
  Strongly Coupled Gauge Theories: Challenges and Perspectives}},\ }\href
  {https://doi.org/10.1140/epjc/s10052-014-2981-5} {\bibfield  {journal}
  {\bibinfo  {journal} {Eur. Phys. J. C}\ }\textbf {\bibinfo {volume} {74}},\
  \bibinfo {pages} {2981} (\bibinfo {year} {2014})},\ \Eprint
  {https://arxiv.org/abs/1404.3723} {arXiv:1404.3723 [hep-ph]} \BibitemShut
  {NoStop}%
\bibitem [{\citenamefont {de~Forcrand}\ and\ \citenamefont
  {Fromm}(2010)}]{deForcrand:2009dh}%
  \BibitemOpen
  \bibfield  {author} {\bibinfo {author} {\bibfnamefont {P.}~\bibnamefont
  {de~Forcrand}}\ and\ \bibinfo {author} {\bibfnamefont {M.}~\bibnamefont
  {Fromm}},\ }\bibfield  {title} {\bibinfo {title} {{Nuclear Physics from
  lattice QCD at strong coupling}},\ }\href
  {https://doi.org/10.1103/PhysRevLett.104.112005} {\bibfield  {journal}
  {\bibinfo  {journal} {Phys. Rev. Lett.}\ }\textbf {\bibinfo {volume} {104}},\
  \bibinfo {pages} {112005} (\bibinfo {year} {2010})},\ \Eprint
  {https://arxiv.org/abs/0907.1915} {arXiv:0907.1915 [hep-lat]} \BibitemShut
  {NoStop}%
\bibitem [{\citenamefont {Kim}\ \emph {et~al.}(2023)\citenamefont {Kim},
  \citenamefont {Pattanaik},\ and\ \citenamefont {Unger}}]{Kim:2023dnq}%
  \BibitemOpen
  \bibfield  {author} {\bibinfo {author} {\bibfnamefont {J.}~\bibnamefont
  {Kim}}, \bibinfo {author} {\bibfnamefont {P.}~\bibnamefont {Pattanaik}},\
  and\ \bibinfo {author} {\bibfnamefont {W.}~\bibnamefont {Unger}},\ }\bibfield
   {title} {\bibinfo {title} {{Nuclear liquid-gas transition in the strong
  coupling regime of lattice QCD}},\ }\href
  {https://doi.org/10.1103/PhysRevD.107.094514} {\bibfield  {journal} {\bibinfo
   {journal} {Phys. Rev. D}\ }\textbf {\bibinfo {volume} {107}},\ \bibinfo
  {pages} {094514} (\bibinfo {year} {2023})},\ \Eprint
  {https://arxiv.org/abs/2303.01467} {arXiv:2303.01467 [hep-lat]} \BibitemShut
  {NoStop}%
\bibitem [{\citenamefont {Holt}\ \emph {et~al.}(2013)\citenamefont {Holt},
  \citenamefont {Kaiser},\ and\ \citenamefont {Weise}}]{Holt:2013fwa}%
  \BibitemOpen
  \bibfield  {author} {\bibinfo {author} {\bibfnamefont {J.~W.}\ \bibnamefont
  {Holt}}, \bibinfo {author} {\bibfnamefont {N.}~\bibnamefont {Kaiser}},\ and\
  \bibinfo {author} {\bibfnamefont {W.}~\bibnamefont {Weise}},\ }\bibfield
  {title} {\bibinfo {title} {{Nuclear chiral dynamics and thermodynamics}},\
  }\href {https://doi.org/10.1016/j.ppnp.2013.08.001} {\bibfield  {journal}
  {\bibinfo  {journal} {Prog. Part. Nucl. Phys.}\ }\textbf {\bibinfo {volume}
  {73}},\ \bibinfo {pages} {35} (\bibinfo {year} {2013})},\ \Eprint
  {https://arxiv.org/abs/1304.6350} {arXiv:1304.6350 [nucl-th]} \BibitemShut
  {NoStop}%
\bibitem [{\citenamefont {Drischler}\ \emph {et~al.}(2019)\citenamefont
  {Drischler}, \citenamefont {Hebeler},\ and\ \citenamefont
  {Schwenk}}]{Drischler:2017wtt}%
  \BibitemOpen
  \bibfield  {author} {\bibinfo {author} {\bibfnamefont {C.}~\bibnamefont
  {Drischler}}, \bibinfo {author} {\bibfnamefont {K.}~\bibnamefont {Hebeler}},\
  and\ \bibinfo {author} {\bibfnamefont {A.}~\bibnamefont {Schwenk}},\
  }\bibfield  {title} {\bibinfo {title} {{Chiral interactions up to
  next-to-next-to-next-to-leading order and nuclear saturation}},\ }\href
  {https://doi.org/10.1103/PhysRevLett.122.042501} {\bibfield  {journal}
  {\bibinfo  {journal} {Phys. Rev. Lett.}\ }\textbf {\bibinfo {volume} {122}},\
  \bibinfo {pages} {042501} (\bibinfo {year} {2019})},\ \Eprint
  {https://arxiv.org/abs/1710.08220} {arXiv:1710.08220 [nucl-th]} \BibitemShut
  {NoStop}%
\bibitem [{\citenamefont {Drischler}\ \emph {et~al.}(2021)\citenamefont
  {Drischler}, \citenamefont {Han}, \citenamefont {Lattimer}, \citenamefont
  {Prakash}, \citenamefont {Reddy},\ and\ \citenamefont
  {Zhao}}]{Drischler:2020fvz}%
  \BibitemOpen
  \bibfield  {author} {\bibinfo {author} {\bibfnamefont {C.}~\bibnamefont
  {Drischler}}, \bibinfo {author} {\bibfnamefont {S.}~\bibnamefont {Han}},
  \bibinfo {author} {\bibfnamefont {J.~M.}\ \bibnamefont {Lattimer}}, \bibinfo
  {author} {\bibfnamefont {M.}~\bibnamefont {Prakash}}, \bibinfo {author}
  {\bibfnamefont {S.}~\bibnamefont {Reddy}},\ and\ \bibinfo {author}
  {\bibfnamefont {T.}~\bibnamefont {Zhao}},\ }\bibfield  {title} {\bibinfo
  {title} {{Limiting masses and radii of neutron stars and their
  implications}},\ }\href {https://doi.org/10.1103/PhysRevC.103.045808}
  {\bibfield  {journal} {\bibinfo  {journal} {Phys. Rev. C}\ }\textbf {\bibinfo
  {volume} {103}},\ \bibinfo {pages} {045808} (\bibinfo {year} {2021})},\
  \Eprint {https://arxiv.org/abs/2009.06441} {arXiv:2009.06441 [nucl-th]}
  \BibitemShut {NoStop}%
\bibitem [{\citenamefont {Leonhardt}\ \emph {et~al.}(2020)\citenamefont
  {Leonhardt}, \citenamefont {Pospiech}, \citenamefont {Schallmo},
  \citenamefont {Braun}, \citenamefont {Drischler}, \citenamefont {Hebeler},\
  and\ \citenamefont {Schwenk}}]{Leonhardt:2019fua}%
  \BibitemOpen
  \bibfield  {author} {\bibinfo {author} {\bibfnamefont {M.}~\bibnamefont
  {Leonhardt}}, \bibinfo {author} {\bibfnamefont {M.}~\bibnamefont {Pospiech}},
  \bibinfo {author} {\bibfnamefont {B.}~\bibnamefont {Schallmo}}, \bibinfo
  {author} {\bibfnamefont {J.}~\bibnamefont {Braun}}, \bibinfo {author}
  {\bibfnamefont {C.}~\bibnamefont {Drischler}}, \bibinfo {author}
  {\bibfnamefont {K.}~\bibnamefont {Hebeler}},\ and\ \bibinfo {author}
  {\bibfnamefont {A.}~\bibnamefont {Schwenk}},\ }\bibfield  {title} {\bibinfo
  {title} {{Symmetric nuclear matter from the strong interaction}},\ }\href
  {https://doi.org/10.1103/PhysRevLett.125.142502} {\bibfield  {journal}
  {\bibinfo  {journal} {Phys. Rev. Lett.}\ }\textbf {\bibinfo {volume} {125}},\
  \bibinfo {pages} {142502} (\bibinfo {year} {2020})},\ \Eprint
  {https://arxiv.org/abs/1907.05814} {arXiv:1907.05814 [nucl-th]} \BibitemShut
  {NoStop}%
\bibitem [{\citenamefont {Mitter}\ \emph {et~al.}(2015)\citenamefont {Mitter},
  \citenamefont {Pawlowski},\ and\ \citenamefont
  {Strodthoff}}]{Mitter:2014wpa}%
  \BibitemOpen
  \bibfield  {author} {\bibinfo {author} {\bibfnamefont {M.}~\bibnamefont
  {Mitter}}, \bibinfo {author} {\bibfnamefont {J.~M.}\ \bibnamefont
  {Pawlowski}},\ and\ \bibinfo {author} {\bibfnamefont {N.}~\bibnamefont
  {Strodthoff}},\ }\bibfield  {title} {\bibinfo {title} {{Chiral symmetry
  breaking in continuum QCD}},\ }\href
  {https://doi.org/10.1103/PhysRevD.91.054035} {\bibfield  {journal} {\bibinfo
  {journal} {Phys. Rev.}\ }\textbf {\bibinfo {volume} {D91}},\ \bibinfo {pages}
  {054035} (\bibinfo {year} {2015})},\ \Eprint
  {https://arxiv.org/abs/1411.7978} {arXiv:1411.7978 [hep-ph]} \BibitemShut
  {NoStop}%
\bibitem [{\citenamefont {Rennecke}(2015)}]{Rennecke:2015eba}%
  \BibitemOpen
  \bibfield  {author} {\bibinfo {author} {\bibfnamefont {F.}~\bibnamefont
  {Rennecke}},\ }\bibfield  {title} {\bibinfo {title} {{Vacuum structure of
  vector mesons in QCD}},\ }\href {https://doi.org/10.1103/PhysRevD.92.076012}
  {\bibfield  {journal} {\bibinfo  {journal} {Phys. Rev.}\ }\textbf {\bibinfo
  {volume} {D92}},\ \bibinfo {pages} {076012} (\bibinfo {year} {2015})},\
  \Eprint {https://arxiv.org/abs/1504.03585} {arXiv:1504.03585 [hep-ph]}
  \BibitemShut {NoStop}%
\bibitem [{\citenamefont {Cyrol}\ \emph
  {et~al.}(2018{\natexlab{a}})\citenamefont {Cyrol}, \citenamefont {Mitter},
  \citenamefont {Pawlowski},\ and\ \citenamefont {Strodthoff}}]{Cyrol:2017ewj}%
  \BibitemOpen
  \bibfield  {author} {\bibinfo {author} {\bibfnamefont {A.~K.}\ \bibnamefont
  {Cyrol}}, \bibinfo {author} {\bibfnamefont {M.}~\bibnamefont {Mitter}},
  \bibinfo {author} {\bibfnamefont {J.~M.}\ \bibnamefont {Pawlowski}},\ and\
  \bibinfo {author} {\bibfnamefont {N.}~\bibnamefont {Strodthoff}},\ }\bibfield
   {title} {\bibinfo {title} {{Nonperturbative quark, gluon, and meson
  correlators of unquenched QCD}},\ }\href
  {https://doi.org/10.1103/PhysRevD.97.054006} {\bibfield  {journal} {\bibinfo
  {journal} {Phys. Rev.}\ }\textbf {\bibinfo {volume} {D97}},\ \bibinfo {pages}
  {054006} (\bibinfo {year} {2018}{\natexlab{a}})},\ \Eprint
  {https://arxiv.org/abs/1706.06326} {arXiv:1706.06326 [hep-ph]} \BibitemShut
  {NoStop}%
\bibitem [{\citenamefont {Braun}\ \emph {et~al.}(2016)\citenamefont {Braun},
  \citenamefont {Fister}, \citenamefont {Pawlowski},\ and\ \citenamefont
  {Rennecke}}]{Braun:2014ata}%
  \BibitemOpen
  \bibfield  {author} {\bibinfo {author} {\bibfnamefont {J.}~\bibnamefont
  {Braun}}, \bibinfo {author} {\bibfnamefont {L.}~\bibnamefont {Fister}},
  \bibinfo {author} {\bibfnamefont {J.~M.}\ \bibnamefont {Pawlowski}},\ and\
  \bibinfo {author} {\bibfnamefont {F.}~\bibnamefont {Rennecke}},\ }\bibfield
  {title} {\bibinfo {title} {{From Quarks and Gluons to Hadrons: Chiral
  Symmetry Breaking in Dynamical QCD}},\ }\href
  {https://doi.org/10.1103/PhysRevD.94.034016} {\bibfield  {journal} {\bibinfo
  {journal} {Phys. Rev.}\ }\textbf {\bibinfo {volume} {D94}},\ \bibinfo {pages}
  {034016} (\bibinfo {year} {2016})},\ \Eprint
  {https://arxiv.org/abs/1412.1045} {arXiv:1412.1045 [hep-ph]} \BibitemShut
  {NoStop}%
\bibitem [{\citenamefont {Fu}\ \emph {et~al.}(2020)\citenamefont {Fu},
  \citenamefont {Pawlowski},\ and\ \citenamefont {Rennecke}}]{Fu:2019hdw}%
  \BibitemOpen
  \bibfield  {author} {\bibinfo {author} {\bibfnamefont {W.-j.}\ \bibnamefont
  {Fu}}, \bibinfo {author} {\bibfnamefont {J.~M.}\ \bibnamefont {Pawlowski}},\
  and\ \bibinfo {author} {\bibfnamefont {F.}~\bibnamefont {Rennecke}},\
  }\bibfield  {title} {\bibinfo {title} {{QCD phase structure at finite
  temperature and density}},\ }\href
  {https://doi.org/10.1103/PhysRevD.101.054032} {\bibfield  {journal} {\bibinfo
   {journal} {Phys. Rev. D}\ }\textbf {\bibinfo {volume} {101}},\ \bibinfo
  {pages} {054032} (\bibinfo {year} {2020})},\ \Eprint
  {https://arxiv.org/abs/1909.02991} {arXiv:1909.02991 [hep-ph]} \BibitemShut
  {NoStop}%
\bibitem [{\citenamefont {Fukushima}\ \emph {et~al.}(2022)\citenamefont
  {Fukushima}, \citenamefont {Pawlowski},\ and\ \citenamefont
  {Strodthoff}}]{Fukushima:2021ctq}%
  \BibitemOpen
  \bibfield  {author} {\bibinfo {author} {\bibfnamefont {K.}~\bibnamefont
  {Fukushima}}, \bibinfo {author} {\bibfnamefont {J.~M.}\ \bibnamefont
  {Pawlowski}},\ and\ \bibinfo {author} {\bibfnamefont {N.}~\bibnamefont
  {Strodthoff}},\ }\bibfield  {title} {\bibinfo {title} {{Emergent hadrons and
  diquarks}},\ }\href {https://doi.org/10.1016/j.aop.2022.169106} {\bibfield
  {journal} {\bibinfo  {journal} {Annals Phys.}\ }\textbf {\bibinfo {volume}
  {446}},\ \bibinfo {pages} {169106} (\bibinfo {year} {2022})},\ \Eprint
  {https://arxiv.org/abs/2103.01129} {arXiv:2103.01129 [hep-ph]} \BibitemShut
  {NoStop}%
\bibitem [{\citenamefont {Serot}\ and\ \citenamefont
  {Walecka}(1997)}]{Serot:1997xg}%
  \BibitemOpen
  \bibfield  {author} {\bibinfo {author} {\bibfnamefont {B.~D.}\ \bibnamefont
  {Serot}}\ and\ \bibinfo {author} {\bibfnamefont {J.~D.}\ \bibnamefont
  {Walecka}},\ }\bibfield  {title} {\bibinfo {title} {{Recent progress in
  quantum hadrodynamics}},\ }\href {https://doi.org/10.1142/S0218301397000299}
  {\bibfield  {journal} {\bibinfo  {journal} {Int. J. Mod. Phys. E}\ }\textbf
  {\bibinfo {volume} {6}},\ \bibinfo {pages} {515} (\bibinfo {year} {1997})},\
  \Eprint {https://arxiv.org/abs/nucl-th/9701058} {arXiv:nucl-th/9701058}
  \BibitemShut {NoStop}%
\bibitem [{\citenamefont {Shen}\ \emph {et~al.}(2019)\citenamefont {Shen},
  \citenamefont {Liang}, \citenamefont {Long}, \citenamefont {Meng},\ and\
  \citenamefont {Ring}}]{Shen:2019dls}%
  \BibitemOpen
  \bibfield  {author} {\bibinfo {author} {\bibfnamefont {S.}~\bibnamefont
  {Shen}}, \bibinfo {author} {\bibfnamefont {H.}~\bibnamefont {Liang}},
  \bibinfo {author} {\bibfnamefont {W.~H.}\ \bibnamefont {Long}}, \bibinfo
  {author} {\bibfnamefont {J.}~\bibnamefont {Meng}},\ and\ \bibinfo {author}
  {\bibfnamefont {P.}~\bibnamefont {Ring}},\ }\bibfield  {title} {\bibinfo
  {title} {{Towards an $ab initio$ covariant density functional theory for
  nuclear structure}},\ }\href {https://doi.org/10.1016/j.ppnp.2019.103713}
  {\bibfield  {journal} {\bibinfo  {journal} {Prog. Part. Nucl. Phys.}\
  }\textbf {\bibinfo {volume} {109}},\ \bibinfo {pages} {103713} (\bibinfo
  {year} {2019})},\ \Eprint {https://arxiv.org/abs/1904.04977}
  {arXiv:1904.04977 [nucl-th]} \BibitemShut {NoStop}%
\bibitem [{\citenamefont {Gies}\ and\ \citenamefont
  {Wetterich}(2004)}]{Gies:2002hq}%
  \BibitemOpen
  \bibfield  {author} {\bibinfo {author} {\bibfnamefont {H.}~\bibnamefont
  {Gies}}\ and\ \bibinfo {author} {\bibfnamefont {C.}~\bibnamefont
  {Wetterich}},\ }\bibfield  {title} {\bibinfo {title} {{Universality of
  spontaneous chiral symmetry breaking in gauge theories}},\ }\href
  {https://doi.org/10.1103/PhysRevD.69.025001} {\bibfield  {journal} {\bibinfo
  {journal} {Phys. Rev.}\ }\textbf {\bibinfo {volume} {D69}},\ \bibinfo {pages}
  {025001} (\bibinfo {year} {2004})},\ \Eprint
  {https://arxiv.org/abs/hep-th/0209183} {arXiv:hep-th/0209183 [hep-th]}
  \BibitemShut {NoStop}%
\bibitem [{\citenamefont {Braun}(2009)}]{Braun:2009ewx}%
  \BibitemOpen
  \bibfield  {author} {\bibinfo {author} {\bibfnamefont {J.}~\bibnamefont
  {Braun}},\ }\bibfield  {title} {\bibinfo {title} {{The QCD Phase Boundary
  from Quark-Gluon Dynamics}},\ }\href
  {https://doi.org/10.1140/epjc/s10052-009-1136-6} {\bibfield  {journal}
  {\bibinfo  {journal} {Eur. Phys. J. C}\ }\textbf {\bibinfo {volume} {64}},\
  \bibinfo {pages} {459} (\bibinfo {year} {2009})},\ \Eprint
  {https://arxiv.org/abs/0810.1727} {arXiv:0810.1727 [hep-ph]} \BibitemShut
  {NoStop}%
\bibitem [{\citenamefont {Dupuis}\ \emph {et~al.}(2021)\citenamefont {Dupuis},
  \citenamefont {Canet}, \citenamefont {Eichhorn}, \citenamefont {Metzner},
  \citenamefont {Pawlowski}, \citenamefont {Tissier},\ and\ \citenamefont
  {Wschebor}}]{Dupuis:2020fhh}%
  \BibitemOpen
  \bibfield  {author} {\bibinfo {author} {\bibfnamefont {N.}~\bibnamefont
  {Dupuis}}, \bibinfo {author} {\bibfnamefont {L.}~\bibnamefont {Canet}},
  \bibinfo {author} {\bibfnamefont {A.}~\bibnamefont {Eichhorn}}, \bibinfo
  {author} {\bibfnamefont {W.}~\bibnamefont {Metzner}}, \bibinfo {author}
  {\bibfnamefont {J.~M.}\ \bibnamefont {Pawlowski}}, \bibinfo {author}
  {\bibfnamefont {M.}~\bibnamefont {Tissier}},\ and\ \bibinfo {author}
  {\bibfnamefont {N.}~\bibnamefont {Wschebor}},\ }\bibfield  {title} {\bibinfo
  {title} {{The nonperturbative functional renormalization group and its
  applications}},\ }\href {https://doi.org/10.1016/j.physrep.2021.01.001}
  {\bibfield  {journal} {\bibinfo  {journal} {Phys. Rept.}\ }\textbf {\bibinfo
  {volume} {910}},\ \bibinfo {pages} {1} (\bibinfo {year} {2021})},\ \Eprint
  {https://arxiv.org/abs/2006.04853} {arXiv:2006.04853 [cond-mat.stat-mech]}
  \BibitemShut {NoStop}%
\bibitem [{\citenamefont {Fu}(2022)}]{Fu:2022gou}%
  \BibitemOpen
  \bibfield  {author} {\bibinfo {author} {\bibfnamefont {W.-j.}\ \bibnamefont
  {Fu}},\ }\bibfield  {title} {\bibinfo {title} {{QCD at finite temperature and
  density within the fRG approach: an overview}},\ }\href
  {https://doi.org/10.1088/1572-9494/ac86be} {\bibfield  {journal} {\bibinfo
  {journal} {Commun. Theor. Phys.}\ }\textbf {\bibinfo {volume} {74}},\
  \bibinfo {pages} {097304} (\bibinfo {year} {2022})},\ \Eprint
  {https://arxiv.org/abs/2205.00468} {arXiv:2205.00468 [hep-ph]} \BibitemShut
  {NoStop}%
\bibitem [{\citenamefont {Jung}\ \emph {et~al.}(2017)\citenamefont {Jung},
  \citenamefont {Rennecke}, \citenamefont {Tripolt}, \citenamefont {von
  Smekal},\ and\ \citenamefont {Wambach}}]{Jung:2016yxl}%
  \BibitemOpen
  \bibfield  {author} {\bibinfo {author} {\bibfnamefont {C.}~\bibnamefont
  {Jung}}, \bibinfo {author} {\bibfnamefont {F.}~\bibnamefont {Rennecke}},
  \bibinfo {author} {\bibfnamefont {R.-A.}\ \bibnamefont {Tripolt}}, \bibinfo
  {author} {\bibfnamefont {L.}~\bibnamefont {von Smekal}},\ and\ \bibinfo
  {author} {\bibfnamefont {J.}~\bibnamefont {Wambach}},\ }\bibfield  {title}
  {\bibinfo {title} {{In-Medium Spectral Functions of Vector- and Axial-Vector
  Mesons from the Functional Renormalization Group}},\ }\href
  {https://doi.org/10.1103/PhysRevD.95.036020} {\bibfield  {journal} {\bibinfo
  {journal} {Phys. Rev. D}\ }\textbf {\bibinfo {volume} {95}},\ \bibinfo
  {pages} {036020} (\bibinfo {year} {2017})},\ \Eprint
  {https://arxiv.org/abs/1610.08754} {arXiv:1610.08754 [hep-ph]} \BibitemShut
  {NoStop}%
\bibitem [{\citenamefont {Jung}\ and\ \citenamefont {von
  Smekal}(2019)}]{Jung:2019nnr}%
  \BibitemOpen
  \bibfield  {author} {\bibinfo {author} {\bibfnamefont {C.}~\bibnamefont
  {Jung}}\ and\ \bibinfo {author} {\bibfnamefont {L.}~\bibnamefont {von
  Smekal}},\ }\bibfield  {title} {\bibinfo {title} {{Fluctuating vector mesons
  in analytically continued functional RG flow equations}},\ }\href
  {https://doi.org/10.1103/PhysRevD.100.116009} {\bibfield  {journal} {\bibinfo
   {journal} {Phys. Rev. D}\ }\textbf {\bibinfo {volume} {100}},\ \bibinfo
  {pages} {116009} (\bibinfo {year} {2019})},\ \Eprint
  {https://arxiv.org/abs/1909.13712} {arXiv:1909.13712 [hep-ph]} \BibitemShut
  {NoStop}%
\bibitem [{\citenamefont {Cohen}(2003)}]{Cohen:2003kd}%
  \BibitemOpen
  \bibfield  {author} {\bibinfo {author} {\bibfnamefont {T.~D.~.}\ \bibnamefont
  {Cohen}},\ }\bibfield  {title} {\bibinfo {title} {{Functional integrals for
  QCD at nonzero chemical potential and zero density}},\ }\href
  {https://doi.org/10.1103/PhysRevLett.91.222001} {\bibfield  {journal}
  {\bibinfo  {journal} {Phys. Rev. Lett.}\ }\textbf {\bibinfo {volume} {91}},\
  \bibinfo {pages} {222001} (\bibinfo {year} {2003})},\ \Eprint
  {https://arxiv.org/abs/hep-ph/0307089} {arXiv:hep-ph/0307089 [hep-ph]}
  \BibitemShut {NoStop}%
\bibitem [{\citenamefont {van Dalen}\ \emph {et~al.}(2005)\citenamefont {van
  Dalen}, \citenamefont {Fuchs},\ and\ \citenamefont
  {Faessler}}]{vanDalen:2005ns}%
  \BibitemOpen
  \bibfield  {author} {\bibinfo {author} {\bibfnamefont {E.~N.~E.}\
  \bibnamefont {van Dalen}}, \bibinfo {author} {\bibfnamefont {C.}~\bibnamefont
  {Fuchs}},\ and\ \bibinfo {author} {\bibfnamefont {A.}~\bibnamefont
  {Faessler}},\ }\bibfield  {title} {\bibinfo {title} {{Effective nucleon
  masses in symmetric and asymmetric nuclear matter}},\ }\href
  {https://doi.org/10.1103/PhysRevLett.95.022302} {\bibfield  {journal}
  {\bibinfo  {journal} {Phys. Rev. Lett.}\ }\textbf {\bibinfo {volume} {95}},\
  \bibinfo {pages} {022302} (\bibinfo {year} {2005})},\ \Eprint
  {https://arxiv.org/abs/nucl-th/0502064} {arXiv:nucl-th/0502064} \BibitemShut
  {NoStop}%
\bibitem [{\citenamefont {Baldo}\ \emph {et~al.}(2014)\citenamefont {Baldo},
  \citenamefont {Burgio}, \citenamefont {Schulze},\ and\ \citenamefont
  {Taranto}}]{Baldo:2014yja}%
  \BibitemOpen
  \bibfield  {author} {\bibinfo {author} {\bibfnamefont {M.}~\bibnamefont
  {Baldo}}, \bibinfo {author} {\bibfnamefont {G.~F.}\ \bibnamefont {Burgio}},
  \bibinfo {author} {\bibfnamefont {H.~J.}\ \bibnamefont {Schulze}},\ and\
  \bibinfo {author} {\bibfnamefont {G.}~\bibnamefont {Taranto}},\ }\bibfield
  {title} {\bibinfo {title} {{Nucleon effective masses within the
  Brueckner-Hartree-Fock theory: Impact on stellar neutrino emission}},\ }\href
  {https://doi.org/10.1103/PhysRevC.89.048801} {\bibfield  {journal} {\bibinfo
  {journal} {Phys. Rev. C}\ }\textbf {\bibinfo {volume} {89}},\ \bibinfo
  {pages} {048801} (\bibinfo {year} {2014})},\ \Eprint
  {https://arxiv.org/abs/1404.7031} {arXiv:1404.7031 [nucl-th]} \BibitemShut
  {NoStop}%
\bibitem [{\citenamefont {Shang}\ \emph {et~al.}(2020)\citenamefont {Shang},
  \citenamefont {Li}, \citenamefont {Miao}, \citenamefont {Burgio},\ and\
  \citenamefont {Schulze}}]{Shang:2020kfc}%
  \BibitemOpen
  \bibfield  {author} {\bibinfo {author} {\bibfnamefont {X.~L.}\ \bibnamefont
  {Shang}}, \bibinfo {author} {\bibfnamefont {A.}~\bibnamefont {Li}}, \bibinfo
  {author} {\bibfnamefont {Z.~Q.}\ \bibnamefont {Miao}}, \bibinfo {author}
  {\bibfnamefont {G.~F.}\ \bibnamefont {Burgio}},\ and\ \bibinfo {author}
  {\bibfnamefont {H.~J.}\ \bibnamefont {Schulze}},\ }\bibfield  {title}
  {\bibinfo {title} {{Nucleon effective mass in hot dense matter}},\ }\href
  {https://doi.org/10.1103/PhysRevC.101.065801} {\bibfield  {journal} {\bibinfo
   {journal} {Phys. Rev. C}\ }\textbf {\bibinfo {volume} {101}},\ \bibinfo
  {pages} {065801} (\bibinfo {year} {2020})},\ \Eprint
  {https://arxiv.org/abs/2001.03859} {arXiv:2001.03859 [nucl-th]} \BibitemShut
  {NoStop}%
\bibitem [{\citenamefont {Gao}\ \emph {et~al.}(2021)\citenamefont {Gao},
  \citenamefont {Papavassiliou},\ and\ \citenamefont
  {Pawlowski}}]{Gao:2021wun}%
  \BibitemOpen
  \bibfield  {author} {\bibinfo {author} {\bibfnamefont {F.}~\bibnamefont
  {Gao}}, \bibinfo {author} {\bibfnamefont {J.}~\bibnamefont {Papavassiliou}},\
  and\ \bibinfo {author} {\bibfnamefont {J.~M.}\ \bibnamefont {Pawlowski}},\
  }\bibfield  {title} {\bibinfo {title} {{Fully coupled functional equations
  for the quark sector of QCD}},\ }\href
  {https://doi.org/10.1103/PhysRevD.103.094013} {\bibfield  {journal} {\bibinfo
   {journal} {Phys. Rev. D}\ }\textbf {\bibinfo {volume} {103}},\ \bibinfo
  {pages} {094013} (\bibinfo {year} {2021})},\ \Eprint
  {https://arxiv.org/abs/2102.13053} {arXiv:2102.13053 [hep-ph]} \BibitemShut
  {NoStop}%
\bibitem [{\citenamefont {Haas}\ \emph {et~al.}(2014)\citenamefont {Haas},
  \citenamefont {Fister},\ and\ \citenamefont {Pawlowski}}]{Haas:2013hpa}%
  \BibitemOpen
  \bibfield  {author} {\bibinfo {author} {\bibfnamefont {M.}~\bibnamefont
  {Haas}}, \bibinfo {author} {\bibfnamefont {L.}~\bibnamefont {Fister}},\ and\
  \bibinfo {author} {\bibfnamefont {J.~M.}\ \bibnamefont {Pawlowski}},\
  }\bibfield  {title} {\bibinfo {title} {{Gluon spectral functions and
  transport coefficients in Yang--Mills theory}},\ }\href
  {https://doi.org/10.1103/PhysRevD.90.091501} {\bibfield  {journal} {\bibinfo
  {journal} {Phys. Rev.}\ }\textbf {\bibinfo {volume} {D90}},\ \bibinfo {pages}
  {091501} (\bibinfo {year} {2014})},\ \Eprint
  {https://arxiv.org/abs/1308.4960} {arXiv:1308.4960 [hep-ph]} \BibitemShut
  {NoStop}%
\bibitem [{\citenamefont {Ilgenfritz}\ \emph {et~al.}(2018)\citenamefont
  {Ilgenfritz}, \citenamefont {Pawlowski}, \citenamefont {Rothkopf},\ and\
  \citenamefont {Trunin}}]{Ilgenfritz:2017kkp}%
  \BibitemOpen
  \bibfield  {author} {\bibinfo {author} {\bibfnamefont {E.-M.}\ \bibnamefont
  {Ilgenfritz}}, \bibinfo {author} {\bibfnamefont {J.~M.}\ \bibnamefont
  {Pawlowski}}, \bibinfo {author} {\bibfnamefont {A.}~\bibnamefont
  {Rothkopf}},\ and\ \bibinfo {author} {\bibfnamefont {A.}~\bibnamefont
  {Trunin}},\ }\bibfield  {title} {\bibinfo {title} {{Finite temperature gluon
  spectral functions from $N_f=2+1+1$ lattice QCD}},\ }\href
  {https://doi.org/10.1140/epjc/s10052-018-5593-7} {\bibfield  {journal}
  {\bibinfo  {journal} {Eur. Phys. J.}\ }\textbf {\bibinfo {volume} {C78}},\
  \bibinfo {pages} {127} (\bibinfo {year} {2018})},\ \Eprint
  {https://arxiv.org/abs/1701.08610} {arXiv:1701.08610 [hep-lat]} \BibitemShut
  {NoStop}%
\bibitem [{\citenamefont {Cyrol}\ \emph
  {et~al.}(2018{\natexlab{b}})\citenamefont {Cyrol}, \citenamefont {Pawlowski},
  \citenamefont {Rothkopf},\ and\ \citenamefont {Wink}}]{Cyrol:2018xeq}%
  \BibitemOpen
  \bibfield  {author} {\bibinfo {author} {\bibfnamefont {A.~K.}\ \bibnamefont
  {Cyrol}}, \bibinfo {author} {\bibfnamefont {J.~M.}\ \bibnamefont
  {Pawlowski}}, \bibinfo {author} {\bibfnamefont {A.}~\bibnamefont
  {Rothkopf}},\ and\ \bibinfo {author} {\bibfnamefont {N.}~\bibnamefont
  {Wink}},\ }\bibfield  {title} {\bibinfo {title} {{Reconstructing the
  gluon}},\ }\href {https://doi.org/10.21468/SciPostPhys.5.6.065} {\bibfield
  {journal} {\bibinfo  {journal} {SciPost Phys.}\ }\textbf {\bibinfo {volume}
  {5}},\ \bibinfo {pages} {065} (\bibinfo {year} {2018}{\natexlab{b}})},\
  \Eprint {https://arxiv.org/abs/1804.00945} {arXiv:1804.00945 [hep-ph]}
  \BibitemShut {NoStop}%
\bibitem [{\citenamefont {Dudal}\ \emph {et~al.}(2019)\citenamefont {Dudal},
  \citenamefont {Oliveira}, \citenamefont {Roelfs},\ and\ \citenamefont
  {Silva}}]{Dudal:2019gvn}%
  \BibitemOpen
  \bibfield  {author} {\bibinfo {author} {\bibfnamefont {D.}~\bibnamefont
  {Dudal}}, \bibinfo {author} {\bibfnamefont {O.}~\bibnamefont {Oliveira}},
  \bibinfo {author} {\bibfnamefont {M.}~\bibnamefont {Roelfs}},\ and\ \bibinfo
  {author} {\bibfnamefont {P.}~\bibnamefont {Silva}},\ }\bibfield  {title}
  {\bibinfo {title} {{Spectral representation of lattice gluon and ghost
  propagators at zero temperature}},\ }\href@noop {} {\  (\bibinfo {year}
  {2019})},\ \Eprint {https://arxiv.org/abs/1901.05348} {arXiv:1901.05348
  [hep-lat]} \BibitemShut {NoStop}%
\bibitem [{\citenamefont {Li}\ \emph {et~al.}(2019)\citenamefont {Li},
  \citenamefont {Lowdon}, \citenamefont {Oliveira},\ and\ \citenamefont
  {Silva}}]{Li:2019hyv}%
  \BibitemOpen
  \bibfield  {author} {\bibinfo {author} {\bibfnamefont {S.~W.}\ \bibnamefont
  {Li}}, \bibinfo {author} {\bibfnamefont {P.}~\bibnamefont {Lowdon}}, \bibinfo
  {author} {\bibfnamefont {O.}~\bibnamefont {Oliveira}},\ and\ \bibinfo
  {author} {\bibfnamefont {P.~J.}\ \bibnamefont {Silva}},\ }\bibfield  {title}
  {\bibinfo {title} {{The generalised infrared structure of the gluon
  propagator}},\ }\href@noop {} {\  (\bibinfo {year} {2019})},\ \Eprint
  {https://arxiv.org/abs/1907.10073} {arXiv:1907.10073 [hep-th]} \BibitemShut
  {NoStop}%
\bibitem [{\citenamefont {Hayashi}\ and\ \citenamefont
  {Kondo}(2021)}]{Hayashi:2021nnj}%
  \BibitemOpen
  \bibfield  {author} {\bibinfo {author} {\bibfnamefont {Y.}~\bibnamefont
  {Hayashi}}\ and\ \bibinfo {author} {\bibfnamefont {K.-I.}\ \bibnamefont
  {Kondo}},\ }\bibfield  {title} {\bibinfo {title} {{Reconstructing confined
  particles with complex singularities}},\ }\href
  {https://doi.org/10.1103/PhysRevD.103.L111504} {\bibfield  {journal}
  {\bibinfo  {journal} {Phys. Rev. D}\ }\textbf {\bibinfo {volume} {103}},\
  \bibinfo {pages} {L111504} (\bibinfo {year} {2021})},\ \Eprint
  {https://arxiv.org/abs/2103.14322} {arXiv:2103.14322 [hep-th]} \BibitemShut
  {NoStop}%
\bibitem [{\citenamefont {Horak}\ \emph
  {et~al.}(2022{\natexlab{a}})\citenamefont {Horak}, \citenamefont {Pawlowski},
  \citenamefont {Rodr\'\i{}guez-Quintero}, \citenamefont {Turnwald},
  \citenamefont {Urban}, \citenamefont {Wink},\ and\ \citenamefont
  {Zafeiropoulos}}]{Horak:2021syv}%
  \BibitemOpen
  \bibfield  {author} {\bibinfo {author} {\bibfnamefont {J.}~\bibnamefont
  {Horak}}, \bibinfo {author} {\bibfnamefont {J.~M.}\ \bibnamefont
  {Pawlowski}}, \bibinfo {author} {\bibfnamefont {J.}~\bibnamefont
  {Rodr\'\i{}guez-Quintero}}, \bibinfo {author} {\bibfnamefont
  {J.}~\bibnamefont {Turnwald}}, \bibinfo {author} {\bibfnamefont {J.~M.}\
  \bibnamefont {Urban}}, \bibinfo {author} {\bibfnamefont {N.}~\bibnamefont
  {Wink}},\ and\ \bibinfo {author} {\bibfnamefont {S.}~\bibnamefont
  {Zafeiropoulos}},\ }\bibfield  {title} {\bibinfo {title} {{Reconstructing QCD
  spectral functions with Gaussian processes}},\ }\href
  {https://doi.org/10.1103/PhysRevD.105.036014} {\bibfield  {journal} {\bibinfo
   {journal} {Phys. Rev. D}\ }\textbf {\bibinfo {volume} {105}},\ \bibinfo
  {pages} {036014} (\bibinfo {year} {2022}{\natexlab{a}})},\ \Eprint
  {https://arxiv.org/abs/2107.13464} {arXiv:2107.13464 [hep-ph]} \BibitemShut
  {NoStop}%
\bibitem [{\citenamefont {Horak}\ \emph {et~al.}(2023)\citenamefont {Horak},
  \citenamefont {Pawlowski}, \citenamefont {Turnwald}, \citenamefont {Urban},
  \citenamefont {Wink},\ and\ \citenamefont {Zafeiropoulos}}]{Horak:2023xfb}%
  \BibitemOpen
  \bibfield  {author} {\bibinfo {author} {\bibfnamefont {J.}~\bibnamefont
  {Horak}}, \bibinfo {author} {\bibfnamefont {J.~M.}\ \bibnamefont
  {Pawlowski}}, \bibinfo {author} {\bibfnamefont {J.}~\bibnamefont {Turnwald}},
  \bibinfo {author} {\bibfnamefont {J.~M.}\ \bibnamefont {Urban}}, \bibinfo
  {author} {\bibfnamefont {N.}~\bibnamefont {Wink}},\ and\ \bibinfo {author}
  {\bibfnamefont {S.}~\bibnamefont {Zafeiropoulos}},\ }\bibfield  {title}
  {\bibinfo {title} {{Nonperturbative strong coupling at timelike momenta}},\
  }\href {https://doi.org/10.1103/PhysRevD.107.076019} {\bibfield  {journal}
  {\bibinfo  {journal} {Phys. Rev. D}\ }\textbf {\bibinfo {volume} {107}},\
  \bibinfo {pages} {076019} (\bibinfo {year} {2023})},\ \Eprint
  {https://arxiv.org/abs/2301.07785} {arXiv:2301.07785 [hep-ph]} \BibitemShut
  {NoStop}%
\bibitem [{\citenamefont {Karsch}\ and\ \citenamefont
  {Kitazawa}(2009)}]{Karsch:2009tp}%
  \BibitemOpen
  \bibfield  {author} {\bibinfo {author} {\bibfnamefont {F.}~\bibnamefont
  {Karsch}}\ and\ \bibinfo {author} {\bibfnamefont {M.}~\bibnamefont
  {Kitazawa}},\ }\bibfield  {title} {\bibinfo {title} {{Quark propagator at
  finite temperature and finite momentum in quenched lattice QCD}},\ }\href
  {https://doi.org/10.1103/PhysRevD.80.056001} {\bibfield  {journal} {\bibinfo
  {journal} {Phys. Rev. D}\ }\textbf {\bibinfo {volume} {80}},\ \bibinfo
  {pages} {056001} (\bibinfo {year} {2009})},\ \Eprint
  {https://arxiv.org/abs/0906.3941} {arXiv:0906.3941 [hep-lat]} \BibitemShut
  {NoStop}%
\bibitem [{\citenamefont {Mueller}\ \emph {et~al.}(2010)\citenamefont
  {Mueller}, \citenamefont {Fischer},\ and\ \citenamefont
  {Nickel}}]{Mueller:2010ah}%
  \BibitemOpen
  \bibfield  {author} {\bibinfo {author} {\bibfnamefont {J.~A.}\ \bibnamefont
  {Mueller}}, \bibinfo {author} {\bibfnamefont {C.~S.}\ \bibnamefont
  {Fischer}},\ and\ \bibinfo {author} {\bibfnamefont {D.}~\bibnamefont
  {Nickel}},\ }\bibfield  {title} {\bibinfo {title} {{Quark spectral properties
  above Tc from Dyson-Schwinger equations}},\ }\href
  {https://doi.org/10.1140/epjc/s10052-010-1499-8} {\bibfield  {journal}
  {\bibinfo  {journal} {Eur. Phys. J. C}\ }\textbf {\bibinfo {volume} {70}},\
  \bibinfo {pages} {1037} (\bibinfo {year} {2010})},\ \Eprint
  {https://arxiv.org/abs/1009.3762} {arXiv:1009.3762 [hep-ph]} \BibitemShut
  {NoStop}%
\bibitem [{\citenamefont {Qin}\ and\ \citenamefont
  {Rischke}(2013)}]{Qin:2013ufa}%
  \BibitemOpen
  \bibfield  {author} {\bibinfo {author} {\bibfnamefont {S.-x.}\ \bibnamefont
  {Qin}}\ and\ \bibinfo {author} {\bibfnamefont {D.~H.}\ \bibnamefont
  {Rischke}},\ }\bibfield  {title} {\bibinfo {title} {{Quark Spectral Function
  and Deconfinement at Nonzero Temperature}},\ }\href
  {https://doi.org/10.1103/PhysRevD.88.056007} {\bibfield  {journal} {\bibinfo
  {journal} {Phys. Rev. D}\ }\textbf {\bibinfo {volume} {88}},\ \bibinfo
  {pages} {056007} (\bibinfo {year} {2013})},\ \Eprint
  {https://arxiv.org/abs/1304.6547} {arXiv:1304.6547 [nucl-th]} \BibitemShut
  {NoStop}%
\bibitem [{\citenamefont {Fischer}\ \emph {et~al.}(2018)\citenamefont
  {Fischer}, \citenamefont {Pawlowski}, \citenamefont {Rothkopf},\ and\
  \citenamefont {Welzbacher}}]{Fischer:2017kbq}%
  \BibitemOpen
  \bibfield  {author} {\bibinfo {author} {\bibfnamefont {C.~S.}\ \bibnamefont
  {Fischer}}, \bibinfo {author} {\bibfnamefont {J.~M.}\ \bibnamefont
  {Pawlowski}}, \bibinfo {author} {\bibfnamefont {A.}~\bibnamefont
  {Rothkopf}},\ and\ \bibinfo {author} {\bibfnamefont {C.~A.}\ \bibnamefont
  {Welzbacher}},\ }\bibfield  {title} {\bibinfo {title} {{Bayesian analysis of
  quark spectral properties from the Dyson-Schwinger equation}},\ }\href
  {https://doi.org/10.1103/PhysRevD.98.014009} {\bibfield  {journal} {\bibinfo
  {journal} {Phys. Rev. D}\ }\textbf {\bibinfo {volume} {98}},\ \bibinfo
  {pages} {014009} (\bibinfo {year} {2018})},\ \Eprint
  {https://arxiv.org/abs/1705.03207} {arXiv:1705.03207 [hep-ph]} \BibitemShut
  {NoStop}%
\bibitem [{\citenamefont {Kamikado}\ \emph {et~al.}(2014)\citenamefont
  {Kamikado}, \citenamefont {Strodthoff}, \citenamefont {von Smekal},\ and\
  \citenamefont {Wambach}}]{Kamikado:2013sia}%
  \BibitemOpen
  \bibfield  {author} {\bibinfo {author} {\bibfnamefont {K.}~\bibnamefont
  {Kamikado}}, \bibinfo {author} {\bibfnamefont {N.}~\bibnamefont
  {Strodthoff}}, \bibinfo {author} {\bibfnamefont {L.}~\bibnamefont {von
  Smekal}},\ and\ \bibinfo {author} {\bibfnamefont {J.}~\bibnamefont
  {Wambach}},\ }\bibfield  {title} {\bibinfo {title} {{Real-time correlation
  functions in the $O(N)$ model from the functional renormalization group}},\
  }\href {https://doi.org/10.1140/epjc/s10052-014-2806-6} {\bibfield  {journal}
  {\bibinfo  {journal} {Eur. Phys. J.}\ }\textbf {\bibinfo {volume} {C74}},\
  \bibinfo {pages} {2806} (\bibinfo {year} {2014})},\ \Eprint
  {https://arxiv.org/abs/1302.6199} {arXiv:1302.6199 [hep-ph]} \BibitemShut
  {NoStop}%
\bibitem [{\citenamefont {Fischer}\ and\ \citenamefont
  {Huber}(2020)}]{Fischer:2020xnb}%
  \BibitemOpen
  \bibfield  {author} {\bibinfo {author} {\bibfnamefont {C.~S.}\ \bibnamefont
  {Fischer}}\ and\ \bibinfo {author} {\bibfnamefont {M.~Q.}\ \bibnamefont
  {Huber}},\ }\bibfield  {title} {\bibinfo {title} {{Landau gauge Yang-Mills
  propagators in the complex momentum plane}},\ }\href
  {https://doi.org/10.1103/PhysRevD.102.094005} {\bibfield  {journal} {\bibinfo
   {journal} {Phys. Rev. D}\ }\textbf {\bibinfo {volume} {102}},\ \bibinfo
  {pages} {094005} (\bibinfo {year} {2020})},\ \Eprint
  {https://arxiv.org/abs/2007.11505} {arXiv:2007.11505 [hep-ph]} \BibitemShut
  {NoStop}%
\bibitem [{\citenamefont {Pawlowski}\ and\ \citenamefont
  {Strodthoff}(2015)}]{Pawlowski:2015mia}%
  \BibitemOpen
  \bibfield  {author} {\bibinfo {author} {\bibfnamefont {J.~M.}\ \bibnamefont
  {Pawlowski}}\ and\ \bibinfo {author} {\bibfnamefont {N.}~\bibnamefont
  {Strodthoff}},\ }\bibfield  {title} {\bibinfo {title} {{Real time correlation
  functions and the functional renormalization group}},\ }\href
  {https://doi.org/10.1103/PhysRevD.92.094009} {\bibfield  {journal} {\bibinfo
  {journal} {Phys. Rev.}\ }\textbf {\bibinfo {volume} {D92}},\ \bibinfo {pages}
  {094009} (\bibinfo {year} {2015})},\ \Eprint
  {https://arxiv.org/abs/1508.01160} {arXiv:1508.01160 [hep-ph]} \BibitemShut
  {NoStop}%
\bibitem [{\citenamefont {Pawlowski}\ \emph {et~al.}(2018)\citenamefont
  {Pawlowski}, \citenamefont {Strodthoff},\ and\ \citenamefont
  {Wink}}]{Pawlowski:2017gxj}%
  \BibitemOpen
  \bibfield  {author} {\bibinfo {author} {\bibfnamefont {J.~M.}\ \bibnamefont
  {Pawlowski}}, \bibinfo {author} {\bibfnamefont {N.}~\bibnamefont
  {Strodthoff}},\ and\ \bibinfo {author} {\bibfnamefont {N.}~\bibnamefont
  {Wink}},\ }\bibfield  {title} {\bibinfo {title} {{Finite temperature spectral
  functions in the O(N)-model}},\ }\href
  {https://doi.org/10.1103/PhysRevD.98.074008} {\bibfield  {journal} {\bibinfo
  {journal} {Phys. Rev.}\ }\textbf {\bibinfo {volume} {D98}},\ \bibinfo {pages}
  {074008} (\bibinfo {year} {2018})},\ \Eprint
  {https://arxiv.org/abs/1711.07444} {arXiv:1711.07444 [hep-th]} \BibitemShut
  {NoStop}%
\bibitem [{\citenamefont {Horak}\ \emph {et~al.}(2020)\citenamefont {Horak},
  \citenamefont {Pawlowski},\ and\ \citenamefont {Wink}}]{Horak:2020eng}%
  \BibitemOpen
  \bibfield  {author} {\bibinfo {author} {\bibfnamefont {J.}~\bibnamefont
  {Horak}}, \bibinfo {author} {\bibfnamefont {J.~M.}\ \bibnamefont
  {Pawlowski}},\ and\ \bibinfo {author} {\bibfnamefont {N.}~\bibnamefont
  {Wink}},\ }\bibfield  {title} {\bibinfo {title} {{Spectral functions in the
  $\phi^4$-theory from the spectral DSE}},\ }\href
  {https://doi.org/10.1103/PhysRevD.102.125016} {\bibfield  {journal} {\bibinfo
   {journal} {Phys. Rev. D}\ }\textbf {\bibinfo {volume} {102}},\ \bibinfo
  {pages} {125016} (\bibinfo {year} {2020})},\ \Eprint
  {https://arxiv.org/abs/2006.09778} {arXiv:2006.09778 [hep-th]} \BibitemShut
  {NoStop}%
\bibitem [{\citenamefont {Horak}\ \emph {et~al.}(2021)\citenamefont {Horak},
  \citenamefont {Papavassiliou}, \citenamefont {Pawlowski},\ and\ \citenamefont
  {Wink}}]{Horak:2021pfr}%
  \BibitemOpen
  \bibfield  {author} {\bibinfo {author} {\bibfnamefont {J.}~\bibnamefont
  {Horak}}, \bibinfo {author} {\bibfnamefont {J.}~\bibnamefont
  {Papavassiliou}}, \bibinfo {author} {\bibfnamefont {J.~M.}\ \bibnamefont
  {Pawlowski}},\ and\ \bibinfo {author} {\bibfnamefont {N.}~\bibnamefont
  {Wink}},\ }\bibfield  {title} {\bibinfo {title} {{Ghost spectral function
  from the spectral Dyson-Schwinger equation}},\ }\bibfield  {journal}
  {\bibinfo  {journal} {Phys. Rev. D}\ }\textbf {\bibinfo {volume} {104}},\
  \href {https://doi.org/10.1103/PhysRevD.104.074017}
  {10.1103/PhysRevD.104.074017} (\bibinfo {year} {2021}),\ \Eprint
  {https://arxiv.org/abs/2103.16175} {arXiv:2103.16175 [hep-th]} \BibitemShut
  {NoStop}%
\bibitem [{\citenamefont {Horak}\ \emph
  {et~al.}(2022{\natexlab{b}})\citenamefont {Horak}, \citenamefont
  {Pawlowski},\ and\ \citenamefont {Wink}}]{Horak:2022myj}%
  \BibitemOpen
  \bibfield  {author} {\bibinfo {author} {\bibfnamefont {J.}~\bibnamefont
  {Horak}}, \bibinfo {author} {\bibfnamefont {J.~M.}\ \bibnamefont
  {Pawlowski}},\ and\ \bibinfo {author} {\bibfnamefont {N.}~\bibnamefont
  {Wink}},\ }\bibfield  {title} {\bibinfo {title} {{On the complex structure of
  Yang-Mills theory}},\ }\href@noop {} {\  (\bibinfo {year}
  {2022}{\natexlab{b}})},\ \Eprint {https://arxiv.org/abs/2202.09333}
  {arXiv:2202.09333 [hep-th]} \BibitemShut {NoStop}%
\bibitem [{\citenamefont {Horak}\ \emph
  {et~al.}(2022{\natexlab{c}})\citenamefont {Horak}, \citenamefont
  {Pawlowski},\ and\ \citenamefont {Wink}}]{Horak:2022aza}%
  \BibitemOpen
  \bibfield  {author} {\bibinfo {author} {\bibfnamefont {J.}~\bibnamefont
  {Horak}}, \bibinfo {author} {\bibfnamefont {J.~M.}\ \bibnamefont
  {Pawlowski}},\ and\ \bibinfo {author} {\bibfnamefont {N.}~\bibnamefont
  {Wink}},\ }\bibfield  {title} {\bibinfo {title} {{On the quark spectral
  function in QCD}},\ }\href@noop {} {\  (\bibinfo {year}
  {2022}{\natexlab{c}})},\ \Eprint {https://arxiv.org/abs/2210.07597}
  {arXiv:2210.07597 [hep-ph]} \BibitemShut {NoStop}%
\bibitem [{\citenamefont {Windisch}(2017)}]{Windisch:2016iud}%
  \BibitemOpen
  \bibfield  {author} {\bibinfo {author} {\bibfnamefont {A.}~\bibnamefont
  {Windisch}},\ }\bibfield  {title} {\bibinfo {title} {{Analytic properties of
  the quark propagator from an effective infrared interaction model}},\ }\href
  {https://doi.org/10.1103/PhysRevC.95.045204} {\bibfield  {journal} {\bibinfo
  {journal} {Phys. Rev. C}\ }\textbf {\bibinfo {volume} {95}},\ \bibinfo
  {pages} {045204} (\bibinfo {year} {2017})},\ \Eprint
  {https://arxiv.org/abs/1612.06002} {arXiv:1612.06002 [hep-ph]} \BibitemShut
  {NoStop}%
\end{thebibliography}%

\end{document}